\documentclass[letter,11pt]{article}
 
\usepackage{jheppub}

\usepackage{graphicx}
\usepackage{amsmath}
\usepackage{slashed}
\usepackage{enumerate}

\newcommand{\nocontentsline}[3]{}
\newcommand{\tocless}[2]{\bgroup\let\addcontentsline=\nocontentsline#1{#2}\egroup}

\newcommand{\nn}{\nonumber}

\newcommand{\bea}{\begin{eqnarray}}
\newcommand{\eea}{\end{eqnarray}}

\def\ln{\textrm{ln}}

\def\nn{\nonumber}

\allowdisplaybreaks[2]

\usepackage{marginnote}

\title{Radiative corrections for factorized jet observables in heavy ion collisions}
\author{Varun Vaidya}
\affiliation{Department of Physics,  University of South Dakota, Vermillion, SD~57069, U.S.A.}

\emailAdd{Varun.Vaidya@usd.edu}

\abstract{I look at the renormalization of the medium structure function and a medium induced jet function in a factorized cross section for jet substructure observables in Heavy Ion collisions. This is based on the formalism developed in \cite{Vaidya:2020lih}, which uses an Open quantum system approach combined with the Effective Field Theory(EFT) for forward scattering to derive a factorization formula for jet observables which work as hard probes of a long lived dilute Quark Gluon Plasma(QGP) medium. I show that the universal medium structure function that captures the observable independent physics of the QGP has both rapidity and UV anomalous dimensions that appear due to medium induced Bremsstrahlung. The resulting Renormalization Group(RG) equations correspond to the BFKL equation and the running of the QCD coupling respectively. I present the first results for the numerical impact of resummation using these RG equations on the mean free path of the jet in the medium. I also briefly discuss the prospects of extending this formalism for a short lived dense medium.
}

\begin{document}
\maketitle

\section{Introduction}

It is now widely accepted that heavy ion collisions are the laboratory for the creation and study of the Quark Gluon Plasma medium. The high energy collision of nuclei both at RHIC and the LHC creates sufficiently energetic partons that can escape confinement from color neutral hadrons and give rise to a strongly/weakly coupled soup of deconfined quarks and gluons known as the Quark Gluon Plasma medium. Experimental evidence suggests that this medium which exists for a very short time ($\sim 10$ fm/c) behaves as a near perfect strongly coupled liquid in thermal equilibrium with very low viscosity.  At the microscopic level, we can think of this plasma as consisting of soft partons with typical energy of the order of the temperature(T) of the medium. The stopping nuclear collisions which create the QGP are accompanied by hard interactions which create highly energetic partons (E $\gg$ T) which eventually form jets. These jets then have to traverse through a region of the hot QGP as they evolve and hence they get modified in heavy ion collision, compared with proton-proton collisions, due to the jet-medium interaction.
The modification of a jet in a medium compared to its vacuum evolution can shed light on the properties of the medium, making them useful hard probes of the Quark Gluon Plasma. A phenomenon that has been extensively studied in  literature\cite{Gyulassy:1993hr,Wang:1994fx,Baier:1994bd,Baier:1996kr,Baier:1996sk,Zakharov:1996fv,Zakharov:1997uu,Gyulassy:1999zd,Gyulassy:2000er,Wiedemann:2000za,Guo:2000nz,Wang:2001ifa,Arnold:2002ja,Arnold:2002zm,Salgado:2003gb,Armesto:2003jh,Majumder:2006wi,Majumder:2007zh,Neufeld:2008fi,Neufeld:2009ep} is that of Jet quenching, which entails a suppression of particles with high transverse momenta in the medium. This has also been observed in experiments at both Relativistic Heavy Ion Collider (RHIC) \cite{Arsene:2004fa,Back:2004je,Adams:2005dq,Adcox:2004mh} and Large Hadron Collider (LHC) \cite{Aad:2010bu,Aamodt:2010jd,Chatrchyan:2011sx}. The suppression mechanism happens through the mechanism of energy loss when jets travel through the hot medium. This can happen either through a collision of the energetic partons in the jet with the soft partons of the medium or through medium-induced radiation, but the latter dominates at high energy. The key to understand jet quenching and jet substructure modifications in heavy ion collisions is to understand how the jet interacts with the expanding medium. There has been tremendous theoretical effort to study the jet energy loss mechanism (see Refs.~\cite{Mehtar-Tani:2013pia,Blaizot:2015lma,Qin:2015srf,Cao:2020wlm} for recent reviews). All these frameworks rely on a direct Feynman diagram calculation which is only valid in a perturbative regime and the information about the medium is encoded in the form of a transport co-efficient $\hat q$ which measures the average transverse momentum gained by a parton per unit time. 

Given the multi-scaled nature of the problem and the fact the the QGP medium created is non-perturbative,  in \cite{Vaidya:2020lih}, I proposed an Effective Field Theory(EFT) approach using tools of Open Quantum systems and SCET(Soft Collinear Effective Theory)(\cite{Bauer:2002aj,Bauer:2003mga,Manohar:2006nz,Bauer:2000yr,Bauer:2001ct,Bauer:2002nz}) to address jet propgation in a strongly coupled QGP. There are several advantages of adopting an EFT approach 
\begin{itemize}
\item
It provides an explicit separation of physics at widely separated scales in the form of a factorization formula for the the jet observables. In the context of Heavy Ion collisions, it was shown in \cite{Vaidya:2020lih} that this completely isolates the universal physics of the medium from the kinematic and observable dependent physics of jet evolution.  
\item
It gives a gauge invariant operator definition for the factorized functions. In particular, the physics of the medium was shown to be captured by a correlator of Soft current computed in the background of the medium density matrix. This definition was independent of the details of the medium density matrix, with the assumption of homogeneity of the medium over certain scales. 
\item 
The factorization does not rely on perturbation theory but only on a separation of scales. So the factorization holds even when some of the functions become non-perturbative. Thus, it gives a clear boundary between the perturbatively calculable and non-calculable physics. This is in exact analogy with factorization formulas for ep and pp experiments such as Deep Inelastic scattering(DIS) and Drell-Yan where the universal physics of the proton is defined by a Parton Distribution Function(PDF). This has a precise operator definition and is non-perturbative while being factorized from process dependent perturbative physics. This has the advantage that if some of the factorized functions become non-perturbative, there is a possibility of computing them independently from lattice/quantum computers.
\item 
Lastly, the separation of scales allows us to resum large logarithms in the ratio of these scales via Renormalization group equation for the factorized functions. This is systematically improvable and has already led to precise predictions in pp and ep experiments.
\end{itemize}

The evolution of the jet in the medium usually depends on multiple scales which can be broadly divided into three categories: Kinematic scales such as such as the jet energy, thermal scales of the QGP , namely the temperature T and the Debye screening mass $m_D \sim gT$ and the size or equivalently the temporal extent of the medium. The second class of scales that appear due to dynamical evolution of the system are the mean free path of the jet and the formation time of the jet, which is the time scale over which the hard partons created in the jet go on-shell. The third and final category are the measurement scales imposed on the final state jet.
The hierarchy between these scales can vary widely depending on the details of the experiment. However, in general the jet energy scale(Q) will be a hard scale much larger than all the other scales. For sufficiently high temperature(T $\gg \Lambda_{QCD}$), we can assume $g \ll 1$ and so that $m_D << T$ which is what we will assume for the rest of this paper. In current heavy ion collision experiments, the temperature achieved lies in the range $150 - 500$ MeV, and may not always be a perturbative scale. Thus, a fully weak coupling calculation may not be valid. In this paper, I will stick to the case of a weakly coupled QGP for simplicity. While this will enable me to do a perturbative calculation, it is not a requirement for deriving a factorization of the physics at widely separated scales. For $T \sim \Lambda_{QCD}$, some of the functions in the factorization formula become non-perturbative and would then need to be extracted from lattice/experiment.

  At the same time, as a first simple case we will assume the size of the medium(L) and the mean free path($\lambda_{\text{mfp}}$) to be much larger than all other time scales in the problem. This may not always be a realistic assumption, in particular the formation time of the jet($t_F$) would be expected to be much larger than L in current experiments especially for high energy jets. At the same time, a dense medium will lead to a small value for $\lambda_{\text{mfp}}$ in which case the well known LPM(Landau-Pomeranchuk-Migdal) effect (\cite{CaronHuot:2010bp,Ke:2018jem,Mehtar-Tani:2019ygg,Arnold:2015qya,Arnold:2016kek}) will become important. However, the assumption $t_F \ll L \sim \lambda_{\text{mfp}}$ greatly simplifies the analysis of the system and gives us a taste of how factorization works out. The more involved but phenomenologically relevant case $t_F \geq \lambda_{\text{mfp}} \gg L$ will be considered in a future paper. I will also argue why the more complicated cases will continue to have the same form for factorization, albeit with some modifications in the definitions of the functions so that most of the framework developed for the simpler case can be carried over.

 The EFT that is extensively used for jet studies at high energy pp colliders is Soft-Collinear Effective Theory (SCET) which provides a systematic approach towards dealing with a multi-scale scattering problem. However, in heavy ion collisions, there is the added complication of the presence of the medium whose detailed time evolution cannot be easily described analytically. A way to deal with our ignorance about the microscopic details of the medium is to use an Open quantum systems approach (see \cite{Breuer:2002pc,OQS} for review) which works by tracing over the medium degrees of freedom and working out the effective evolution of the reduced density matrix of the system(jet). 

There are also formulations of SCET (known as SCET$_G$) treating the Glauber gluon, which is a type of mode appearing in forward scattering, as a background field induced by the medium interacting with an energetic jet. By making use of the collinear sector of the corresponding  EFT, this formalism has been used to address the question of jet quenching in the medium \cite{Ovanesyan:2011kn,Chien:2015hda,Ovanesyan:2011xy,Chien:2015vja,Kang:2014xsa}. I will use a modern approach using a new EFT for forward scattering that has been developed recently \cite{Rothstein:2016bsq} which also uses the Glauber mode to write down contact operators between the soft and collinear momentum degrees of freedom which is ideally suited for the situation we want to study. 

 \cite{Vaidya:2020cyi} looked at the transverse momentum spread of a single energetic quark as a function of the time of propagation through the QGP medium. However for a realistic description of the system, we also need to account for the initial hard interaction that creates  the energetic quark which is dressed with radiation from the subsequent parton shower which accounts for any vacuum evolution of the jet along with any medium interactions.

Combining SCET with Open quantum systems, a factorization formula for a illustrative jet substructure observable taking into account the hard interaction and subsequent parton shower was derived in \cite{Vaidya:2020lih} for an  idealized case of a long lived dilute medium. This paper introduced a universal medium structure function for the Quark Gluon Plasma in analogy with a PDF for hadron structure along with a medium induced jet function. Radiative corrections corresponding to elastic collisions of the jet with the medium were calculated in this paper and shown to be UV finite. 

In this paper, I will present the results for radiative corrections from medium induced radiation. I will show, for the first time that these corrections induce rapidity and UV divergences in the medium structure and jet functions which help determine the Renormalization Group(RG) evolution of these functions. The solution for these RG equations helps resum logarithms of widely separated scales which form the dominant corrections compared to those from elastic collisions. I will discuss the numerical impact of resummation on the mean free path of the jet in the medium.

The long term goal is to develop a theoretically robust formalism for calculating jet substructure observables for both light parton and heavy quark jets. For example, the bottom quark jets have been identified as an effective probe of the QGP medium and will be experimentally studied at LHC, as well as by the sPHENIX collaboration at RHIC. There has been recent work on computing jet substructure observables for heavy quark jets in the context of proton-proton collisions \cite{Lee:2019lge,Makris:2018npl}. The objective would then be to compute the same observables in heavy ion collisions and study modifications caused by the medium.

This paper is organized as follows.
In Section~\ref{sec:Fact} I review the details of the observable and factorization formula derived in \cite{Vaidya:2020lih}  and present the corrections from elastic collisions of the jet partons with the medium. In Section \ref{sec:Loop} I detail the one loop corrections from medium induced Brehmstrahlung for the medium structure function and the medium jet function which also gives their Renormalization Group(RG) equations. Section~\ref{sec:Resum} gives analytical results for the solution of RG evolution equations. Section~\ref{sec:Master} provides the solution of the Lindblad type evolution equation which resums multiple incoherent interactions of the jet with the medium with a RG evolved medium kernel. Section~\ref{sec:Num} discusses the numerical impact of RG evolution on the mean free path of the jet in the QGP medium. A summary and analysis of results along with future directions in provided in Section~\ref{sec:Conclusion}.


\section{Factorization for Jet substructure observables}
\label{sec:Fact}
In \cite{Vaidya:2020lih}, I developed an EFT formalism for jet substructure observables in a heavy ion collision environment, writing down a factorization formula for an illustrative observable: The transverse momentum imbalance between groomed dijets along with a cumulative jet mass measurement on each jet. This observable was introduced to allow for a clean measurement while countering the issue of jet selection bias. 
We want to consider final state fat (jet radius R $\sim $ 1) dijet events produced in a heavy ion collision in the background of a QGP medium. The jets are isolated using a suitable jet algorithm such as anti-kT with jet radius $R\sim 1$. We examine the scenario when the hard interaction creating the back to back jets happens at the periphery of the heavy ion collision so that effectively only one jet passes through the medium while the other evolves purely in vacuum. \\
Since, it is hard to give a reliable theory prediction for the distribution of soft hadrons from the cooling QGP medium, we put a grooming on the jets in the form of Soft-Drop algorithm \cite{Larkoski:2014wba} with $\beta=0$. The energy cut-off is chosen sufficiently large to remove all partons at energy T and lower. Given a hard scale Q $\sim 2E_J$, where $E_J$ is the energy of the jet and an energy cut-off, $z_{c}E_J$, we work in the hierarchy 
\bea
Q \sim z_{c}Q \gg T 
\eea
where T is the plasma temperature.  The measurement we wish to impose is the transverse momentum imbalance between the two jets and we want to to give predictions for the regime $q_T \sim T$. We are going to assume a high temperature weak coupling $g \ll 1$ scenario so that the Debye screening mass $m_D$ is parametrically much smaller than the temperature.

While this constrains the radiation that lies outside the groomed jet, we still need to ensure that the radiation that passes grooming and hence forms the jet has a single hard subjet. This can be ensured by putting a cumulative jet mass measurement $e_i$ on each groomed jet with the scaling $e_i \sim  T^2/Q^2 $. 
 
We wish to write down a factorization theorem within Soft Collinear Effective Theory(SCET) which separates out functions depending on their scaling in momentum space with $ \lambda = q_T/Q \sim T/Q \sim \sqrt{e_i}$ as the expansion parameter of our EFT. 
The dominant interaction of the jet with the medium involves forward scattering of the jet in the medium and is mediated by the Glauber mode. Using an open quantum systems approach combined with the EFT for forward scattering developed in \cite{Rothstein:2016bsq}, a factorization formula for the reduced density matrix of the jet was derived in \cite{Vaidya:2020lih}. This factorization was derived under the assumption of a long lived dilute medium meaning
\begin{itemize}
\item
The length of the medium L which is equivalent to the time of propagation of the jet in the medium and the mean free path of the jet are larger than than the formation time of the jet.  
\item
The medium is homogeneous over length scales probed by the jet.
\end{itemize}
These assumptions implied that the dominant contribution to the cross section was when the partons created in the jet went on-shell before interacting with the medium and successive interactions of the jet with the medium were incoherent. 
Upto quadratic order in the Glauber Hamiltonian expansion which is equivalent to a single interaction of the jet with the medium, we can write a factorized formula for the trace over the reduced density matrix of the jet as a function of the time of propagation in the medium(t) with the jet substructure measurement $\mathcal{M}$ described above.
\bea
 &&\text{Tr}[\rho(t) \mathcal{M}]\equiv \Sigma(e_n,\vec{q}_T,t)\nn\\
&=& V\times \Bigg[\Sigma^{(0)}(e_n,\vec{q}_T)+ t |C_G(Q)|^2\Sigma^{(0)}(e_n,\vec{q}_T)\otimes_{q_T}\int d^2k_{\perp}S_G^{AB}(k_{\perp})\mathcal{J}_{n,\text{Med}}^{AB}\Bigg]+O(H_G^3)+..\nn
\label{Rho}
\eea 
V is the 4 d volume factor. $\Sigma^{(0)}(e_n,\vec{q}_T)$ is the vacuum density matrix which captures the evolution of the jet in a vacuum background, which is further factorized in terms of vacuum soft(S) and jet($\mathcal{J}_i^{\perp}$) functions.
\bea
\label{SigmZ}
\Sigma^{(0)}(q_T,e_n,e_{\bar n}) = H(Q,\mu)S(\vec{q}_T;\mu,\nu)\otimes_{q_T}\mathcal{J}^{\perp}_n(e_n,Q,z_{c},\vec{q}_T;\mu,\nu)\otimes_{q_T} \mathcal{J}^{\perp}_{\bar n}(e_{\bar n},Q,z_{c},\vec{q}_T;\mu,\nu)\nn\\
\eea
where $\otimes_{q_T}$ indicates a convolution in $\vec{q}_T$. $\vec{q}_T$ is the transverse momentum imbalance between the two jets (n and $\bar n$) with $e_n$ and $e_{\bar n}$ as the cumulative jet mass measurement respectively.
H(Q) is the hard function which also includes the born level term. The factorized functions are defined as follows 
\bea
  S(\vec{q}_T) =  \frac{1}{N_R}\text{tr}  \langle X_S| \mathcal{T}\Big\{e^{-i\int_0^{\infty}dt' H_S(t')}S_{\bar n}^{\dagger}S_n(0)\Big\}\mathcal{\bar T}\Big\{e^{-i\int_0^{\infty}dt' H_S(t')}S_{n}^{\dagger}S_{\bar n}(0)\Big\}\delta^2(\vec{q}_T-\mathcal{P}_{\perp})|X_S\rangle \nn
\eea
where $S_i$ is a Soft Wilson line in the fundamental representation 
\bea
S_n^{(r)}(x)= \mathcal{P}\exp \Bigg[ig\int_{\-\infty}^{0}ds n\cdot A_s^B(x+sn)T^B_{(r)}\Bigg]
\label{SWilson}
\eea
$H_S$ is the soft Hamiltonian which is equivalent to the full QCD Hamiltonian.
The vacuum quark jet function is defined as 
\bea
\mathcal{J}_n^{\perp}(e,Q) &=& \frac{(2\pi)^3}{N_c} \text{tr}\langle X_n| \mathcal{T}\Big\{e^{-i\int_0^{\infty}dt' H_n(t')}\bar \chi_n(0)\Big\}|0\rangle\nn\\
&& \langle 0|\mathcal{\bar T}\Big\{e^{-i\int_0^{\infty}dt' H_n(t')}\frac{\slashed{\bar n}}{2}\chi_n\Big\}\delta(Q-\mathcal{P}^-)\delta^2(\mathcal{P}^{\perp})\Theta(e_n- \mathcal{E}_{\in n,\text{gr}})\delta^2(\vec{q}_T-\mathcal{P}^{\perp}_{\not\in n,\text{gr}})|X_n\rangle  \nn
\eea
with $\in n,\text{gr}$ refers to collinear partons that pass grooming. $n \equiv (1,0,0,1)$ refers to the direction of the jet.  A similar definition holds for the $\bar n \equiv (1,0,0,-1)$ jet. $H_n$ is the collinear Hamiltonian of SCET.
The interaction with the medium introduces three new functions, namely the Glauber hard co-efficient $C_G(\mu)$, medium structure function $S_G^{AB}$, and the medium jet function $\mathcal{J}^{AB}_{n,\text{Med}}$. This is under the assumption that only the n jet sees the medium. 
The function $C_G(\mu)$ is simply $8\pi \alpha_s(\mu)$ with a natural scale $\mu  \sim Q$.
The medium jet function is the difference of a real and virtual medium jet function written in impact parameter ($\vec{b}$) space which is conjugate to $\vec{q}_T$.
\bea
\label{JetMed}
\mathcal{J}^{AB}_{n,\text{Med}}= \mathcal{J}_{n,R}^{AB}-\mathcal{J}_{n,V}^{AB}
\eea
\bea
 \mathcal{J}_{n,R}^{AB}(e_n, \vec{b}, \vec{k}_{\perp})&=&\frac{1}{\mathcal{J}^{\perp}_n(e_n,b)k_{\perp}^2}\frac{1}{[p_{\tilde X}^-]^2}\int \frac{d^2q_ne^{-i\vec{q}_n\cdot \vec{b}}}{(2\pi)^2}\langle p_{\tilde X}-k, \bar{X}_n|\delta^2(\mathcal{P}_{\perp})\delta(Q-p^-)\bar\chi_n(0)\frac{\slashed{\bar n}}{2}|0\rangle \nn\\
	&\times&  \langle 0|\chi_n(0)|p_{\tilde X}-k, \bar{X}_n\rangle  \langle \tilde X_n|O_n^A(0)|p_{\tilde X}-k\rangle \langle p_{\tilde X}-k|O_n^B(0)|\tilde X_n\rangle \mathcal{M}_{X_n}
\eea
This describes the real interaction of the medium with a single parton of the jet at time. $p_{\tilde X}-k$ is the momenta of the single collinear intermediate parton state which interacts with the medium $\tilde X_n$, while $|X_n \rangle$ is a complete set of collinear final states. There is an implicit integral over the phase space of all states. ($X_n, \tilde X_n$).
$\mathcal{M}_{X_n}$ is the transverse momentum and jet mass measurement imposed on the final state $|X_n, \tilde X_n \rangle$ 
\bea
\mathcal{M} = \Theta(e_n-\mathcal{E}_{\in \text{gr}})\delta^2(\vec{q}_n-\mathcal{P}^{\perp}_{\notin \text{gr}}-\vec{k}_{\perp})
\label{measure}
\eea
\bea 
\label{Jetv}
\mathcal{J}^{AB}_{n,V}(b) &= & \frac{1}{k_{\perp}^2 \mathcal{J}^{\perp}_n(e_n,b)} \frac{1}{[p_{\tilde Y}^-]^2}\int \frac{d^2q_ne^{-i\vec{q}_n\cdot \vec{b}}}{(2\pi)^2} \langle p_{\tilde Y}+k|O_n^A(0)|\tilde Y_n \rangle \langle \tilde Y_n|O_n^B(0)|p_{\tilde Y}+k \rangle \nn\\
&&\langle p_{\tilde Y}+k,\bar {X}_{n}|\chi_n(0)\frac{\slashed{\bar n}}{2}|0\rangle \langle 0|\bar \chi_n(0)\delta^2(\mathcal{P}_{\perp})\delta^2(Q-\mathcal{P}^-)\mathcal{M}_V|p_{\tilde Y}+k,\bar{X}_n\rangle
\eea
This describes the virtual interaction of a jet parton($|\tilde Y_n\rangle $) with the medium, again considering one parton at a time. The measurement $\mathcal{M}_V$ in this case is identical to that in Eq.\ref{measure} but acts on the final states $|X_n\rangle,|\tilde{Y}_n \rangle$. $\mathcal{J}^{\perp}_n(e_n,b)$ is the vacuum jet function in impact parameter space. Both the vacuum and medium jet function are defined in terms of the dressed quark field $\chi_n$.
\bea
\chi_n = W_n^{\dagger}\xi_n, \ \ \ \ W_n(x)= \mathcal{P}\exp\left(ig\int_{\infty}^0ds \bar n \cdot A_n(x+\bar n s)\right)
\eea
The medium jet function also has a gauge invariant quark and gluon collinear current 
\bea
O_n^A = \bar{\chi_n}T^A \frac{\slashed{\bar n}}{2} \chi_n +\frac{i}{2}f^{ACD}\mathcal{B}_{n\perp \mu}^C\frac{\bar n}{2}\cdot(\mathcal{P}+\mathcal{P}^{\dagger})\mathcal{B}_{n\perp}^{D\mu}\nn\\
\mathcal{B}_{n\perp}^{\mu}=\mathcal{B}_{n\perp}^{\mu A}T^A= \frac{1}{g}\Big[W_n^{\dagger}iD_{n\perp}^{\mu}W_n\Big]
\eea

 The medium structure function is defined as 
\bea
\label{MedSoft}
S_G^{AB}(k_{\perp}) = \int \frac{dk^-}{2\pi}\frac{1}{k_{\perp}^2}D^{AB}(k_{\perp},k^-)
\eea
where
\bea
D_>^{AB}(k) = \int d^4x e^{-i k \cdot x} \langle X_S|O_S^{A}(x)\rho O_S^B(0)|X_S\rangle 
\eea
is the correlator of Soft operators in the QGP background density matrix $\rho$. The Soft operator $O_S^A$ is a sum over gauge invariant Quark and Gluon soft currents defined in \cite{Rothstein:2016bsq}
\bea
O_S^{A} =  \bar{\psi}_S^nT^A\frac{\slashed{n}}{2}\psi_S^n+\frac{i}{2}f^{ACD}\mathcal{B}_{S\perp\mu}^C\frac{n}{2}\cdot(\mathcal{P}+\mathcal{P}^{\dagger})\mathcal{B}_{S\perp}^{n\mu D}
\eea
 with 
\bea
\psi_s^n=S_n^{\dagger}\psi_s, \ \ \ \  \mathcal{B}_{S\perp \mu}^{n} = \mathcal{B}_{S\perp \mu}^{n A}T^A = \frac{1}{g}\Big[S_n^{\dagger}iD_{S_{\perp}}^{\mu}S_n\Big]
\eea
defined in terms of the Soft Wilson line (Eq.\ref{SWilson})
The one loop results and the corresponding renormalization group equations for the vacuum soft and jet functions were derived in \cite{Vaidya:2020lih}. For the medium structure function and the medium induced jet function there are two types of radiative corrections at one loop:
\begin{enumerate}
\item
 Elastic collisions of the jet partons with the medium which were also computed in \cite{Vaidya:2020lih}. These corrections are UV finite  but IR($m_D$) sensitive. Hence they do not contribute to the renormalization of these functions.
\item
Medium induced Brehmstrahlung which will be considered in detail in this paper in the next section.
\end{enumerate}


\section{One loop results for medium induced functions}
\label{sec:Loop}

In this section we will look at the complete set of radiative corrections upto Next-to-Leading order for our medium soft and jet function. To compute the result for the Soft function, we need to assume a form for the density matrix of the medium. A relevant choice is a thermal density matrix 
\bea
\rho = \frac{e^{-\beta H_S}}{\text{Tr}\Big[e^{-\beta H_s}\Big]}
\eea
It should be emphasized that the RG equations for this function are independent of the state that is chosen for the medium since they are a property of the operator.
The tree level result for the Soft function along with  partial one loop results for the medium jet function corresponding to elastic collisions with the medium were presented in \cite{Vaidya:2020lih}. Here we will evaluate the one loop corrections for the medium Soft function as well as the remainder of the one loop medium jet function which correspond to medium induced radiation.


\subsection{Medium Jet function} 

The medium jet function is defined in Eq.\ref{JetMed} as the difference between the virtual and real interaction with the medium. The tree level result is 
\bea
\mathcal{J}_{n,\text{Med}}^{AB,(0)}(\vec{k}_{\perp},b,e_n) = \frac{e^{-i\vec{k}_{\perp}\cdot \vec{b}}-1}{k_{\perp}^2} 
\eea
The medium induced jet function encodes the modification of the jet due to the interaction with the medium. The complete set of corrections for this function at one loop involves: 
\begin{itemize}
\item{} Elastic collisions of the jet partons with the medium.
\item{} Medium induced radiation. 
\end{itemize}
The details of the one loop computation for the elastic collisions are computed in \cite{Vaidya:2020lih}.
\bea
\mathcal{J}_{n,\text{Med},\text{Elastic}}^{AB(1)}&=& \frac{\frac{1}{2}\delta^{AB}}{k_{\perp}^2}\left(Q(b) + G(b)\right)
\eea
where the result is written in terms of Quark and Gluon contributions 
\small
\bea
 Q(b)&=&\frac{\alpha_s C_F e^{-i\vec{k}_{\perp}\cdot \vec{b}}}{2 \pi} \Bigg[\int_{z_c}^{1-z_c}dz p_{gq}(z)\Big\{\Theta(z_c-\frac{y}{e_n+y})\ln \left(\frac{-B(z)}{e_n(1-z)}\right)-\Theta(z_c-\frac{e_n}{e_n+y})\ln \frac{B(z)e_n}{yM}\Big\}\nn\\
&+&\Theta( \frac{e_n}{e_n+y}-z_c)\Theta(\frac{y}{e_n+y}-z_c)\Big\{\int_{z_c}^{\frac{e_n}{e_n+y}}dz p_{gq}(z)\ln \left(\frac{-B(z)}{e_n(1-z)}\right)-\int_{\frac{e_n}{e_n+y}}^{1-z_c}dz p_{gq}(z)\ln \frac{B(z)e_n}{yM}\Big\}\Bigg]\nn\\
&+& \frac{\alpha_sC_F}{2\pi}\left(e^{-i\vec{k}_{\perp}\cdot \vec{b}}-1\right)\int_{1-z_c}^{1}p_{gq}(z)\ln \frac{m_D^2b^2e^{2\gamma_E}(1-z)}{4}
\eea
The Gluon operator contribution G(b) is given by the difference $R_g-V_g$, where 
\bea 
R_g&=& \frac{\alpha_s N_c e^{-i\vec{k}_{\perp}\cdot \vec{b}}}{2 \pi} \Bigg[\int_{z_c}^{1-z_c}dz p_{gq}(z)\Big\{\Theta(z_c-\frac{y}{e_n+y})\ln\frac{A(z)}{M}-\Theta(z_c-\frac{e_n}{e_n+y})\ln \left(\frac{-A(z)}{y(1-z)}\right)\Big\}\nn\\
&-&\Theta( \frac{e_n}{e_n+y}-z_c)\Theta(\frac{y}{e_n+y}-z_c)\Big\{\int_{z_c}^{\frac{y}{e_n+y}}dz p_{gq}(z)\ln \left(\frac{-A(z)}{y(1-z)}\right)-\int_{\frac{y}{e_n+y}}^{1-z_c}dz p_{gq}(z)\ln \frac{A(z)}{M}\Big\}\Bigg]\nn\\
&-& \frac{\alpha_sN_c}{2\pi}e^{-i\vec{k}_{\perp}\cdot \vec{b}}\int_{1-z_c}^{1}p_{gq}(z)\ln \frac{m_D^2b^2e^{2\gamma_E}(1-z)}{4}
\eea 
\normalsize
written in terms of
\bea
p_{gq}(z) = \frac{1+ (1-z)^2}{z}, \ \ \  y = \frac{4k_{\perp}^2}{Q^2}, \ \ \ M =\frac{4m_D^2}{Q^2}, \ \ A(z) =e_nz-(1-z)y , \ \ B(z) =zy-e_n(1-z)\nn
\eea 
and $V_g$ is obtained by simply evaluating $R_g$ at $k_{\perp}=0$.
The integrals over z can be done exactly analytically but are not too illuminating and hence we refrain from presenting them here. $k_{\perp}$ is the transverse momentum exchanged with between the jet and the medium. The contributions Q(b) and G(b) go to 0 as $k_{\perp}$ goes to 0 so that it is a purely medium induced result. As we see here, this result in UV finite but is sensitive to the IR scale $m_D$ in the form of single logarithms $\ln Q^2e_n/m_D^2 \sim \ln T^2/m_D^2 \sim \ln 1/g^2$. For the case $g \ll 1$, ideally we would like to resum these logarithms. In principle, this can be achieved by matching the jet function to an EFT at the scale $m_D$, which is something we leave for the future. As we will see, the dominant radiative corrections actually come from medium induced radiation and we will focus on their resummation and numerical analysis in this paper.

We will now compute in detail, the corrections for medium induced Bremsstrahlung. We can separately consider the corrections for $\mathcal{J}^{AB}_{n,R}$ and $\mathcal{J}_{n,V}^{AB}$ and evaluate the result first in transverse momentum $\vec{q}_{Tn}$ space before moving to the impact parameter ($\vec{b}$) space.  We will begin with $\mathcal{J}^{AB}_{n,R}$.


\subsubsection{$\mathcal{J}^{AB}_{n,R}$}
\begin{figure}
  \includegraphics[width=\linewidth]{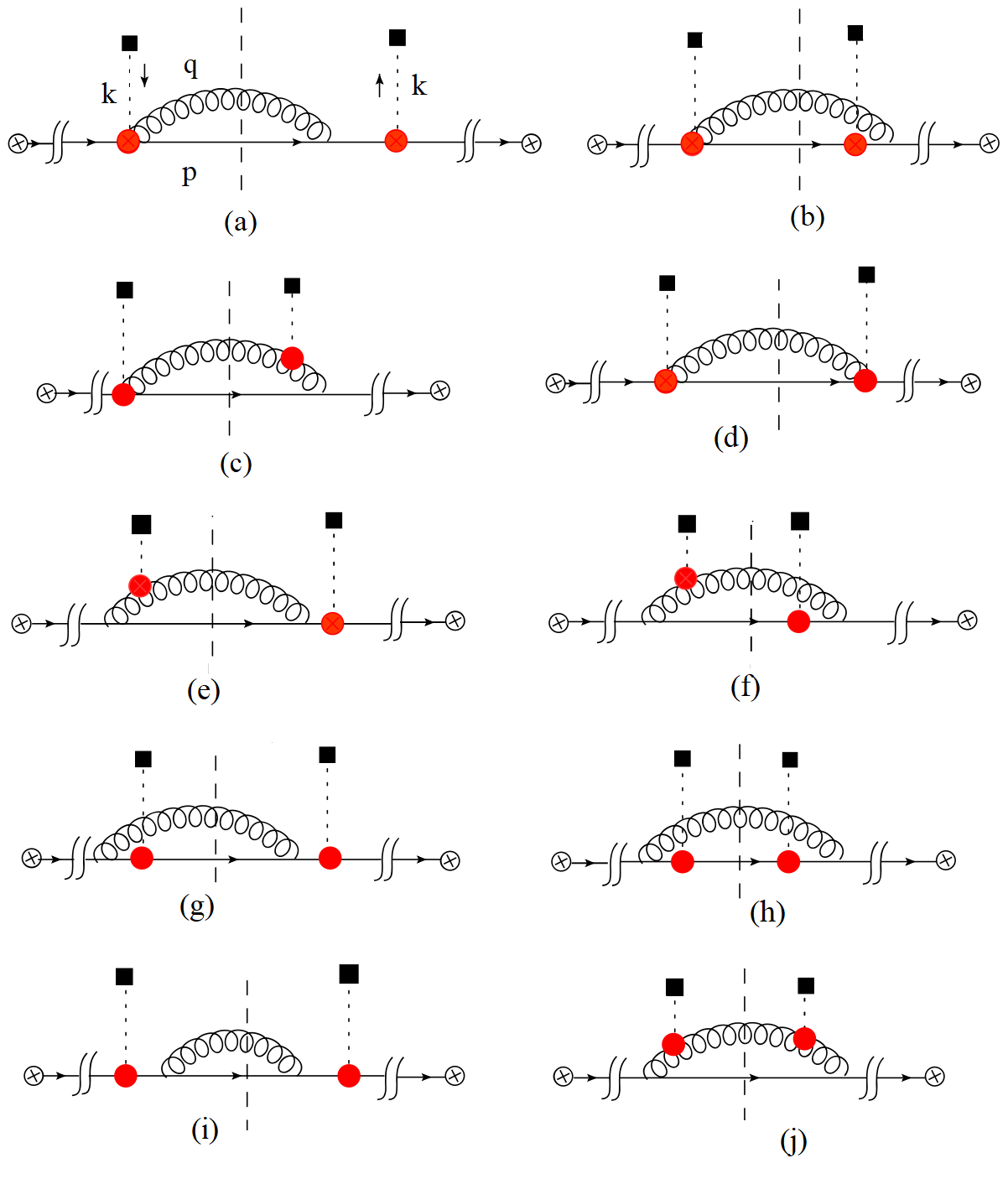}
  \caption{Real gluon emission corrections for $\mathcal{J}^{AB}_{n,R}$. k is the momentum transferred to the jet from the medium. The $\otimes$ vertex refers to the hard vertex which creates a high energy quark that goes on-shell before interacting with the medium. The red dot is the Glauber collinear current and contains both quark and gluon interaction with the medium.}
  \label{Real}
\end{figure}

There are both real and virtual diagrams. The complete set of real diagrams is given in Fig. \ref{Real} and correspond to a Glauber exchange on each side of the cut. The real diagrams are all UV finite due to the imposed measurement on the jet and hence we can work in 4 dimensions.

\begin{itemize}
\item{} Diagram (a) involves a contribution from the Wilson line and an insertion from the collinear Lagrangian. This evaluates to
\small 
\bea
 G_a^R &=& 2N_c\delta^{AB}g^2\nu^{\eta}\int dp^+\int \frac{d^4q}{(2\pi)^{3}}\delta^+(p^+-\frac{(\vec{q}_{\perp}+\vec{k}_{\perp})^2}{Q-q^-})\delta^+(q^2-m_D^2)\frac{Q|q^-|^{-\eta}}{q^-[(p^++q^+)Q-\vec{k}_{\perp}^2]}\mathcal{M}\nn
\eea
\normalsize
where the measurement function is 
\bea
\mathcal{M}&=&\Big[\Theta(Q-q^--Qz_{c})\Theta(q^--Qz_{c})\Theta(e_n- \frac{4}{Q^2}(p+q)^2)\delta^2(\vec{q}_{Tn}-\vec{k}_{\perp})\nn\\
&+& \Theta(Qz_{c}-q^-)\delta^2(\vec{q}_{Tn}-\vec{k}_{\perp}-\vec{q}_{\perp})+\Theta(Qz_{c}-(Q-q^-))\delta^2(\vec{q}_{Tn}+\vec{q}_{\perp}-\vec{k}_{\perp})\Big]\nn
\eea
p is the final state quark momentum while q is the gluon momentum. We then have three contributions from the measurement functions. Among these, only the second one where q fails the grooming condition leads to a divergence while the other contributions are all finite. The divergence in the second term is a rapidity divergence as $q^-\rightarrow 0$ which is regulated by our rapidity regulator $\nu$ and will give the dominant contribution. We will concentrate on extracting this contribution to the anomalous dimension of the medium jet function.

We can isolate the rapidity pole ignoring finite corrections,
\bea
 G_{a,\eta}^R&=&N_c\delta^{AB}g^2(\nu/Q)^{\eta}\Bigg[\int_0^{z_{c}} \frac{dzz^{-1-\eta}}{(2\pi)^{3}}\Bigg]\int d^2q_{\perp}\frac{1}{[(\vec{q}_{\perp}-\vec{k}_{\perp})^2+m_D^2]}\delta^2(\vec{q}_{Tn}-\vec{q}_{\perp})\nn\\
&=& -N_c\delta^{AB}g^2(\nu/[Qz_c])^{\eta}\frac{1}{\eta}\frac{1}{(2\pi)^{3}}\int d^2q_{\perp}\frac{1}{[(\vec{q}_{\perp}-\vec{k}_{\perp})^2+m_D^2]}\delta^2(\vec{q}_{Tn}-\vec{q}_{\perp})
\eea 


\item
The next diagram in shown in Fig. \ref{Real}(b) which is similar to Fig. \ref{Real}(a) and gives us 
\bea
 G_b^R &=& 4\delta^{AB}N_cg^2\nu^{\eta}\int d^4p \delta^+(p^2)\int \frac{d^4q}{(2\pi)^{d-1}}\delta(Q-p^--q^-)\delta^+(q^2-m_D^2)\delta^2(q_{\perp}+p_{\perp}+\vec{k}_{\perp})\nn\\
&\times& \frac{(p^-)^2|q^-|^{-\eta}}{q^-((p+k)^2)}\mathcal{M}
\eea
which once again has a rapidity divergence which gives the same contribution as that from diagram Fig. \ref{Real}(a), which we can isolate, 
\bea
 G_{b,\eta}^R&=&-N_c\delta^{AB}g^2(\nu/[Qz_c])^{\eta}\frac{1}{\eta}\frac{1}{(2\pi)^{3}}\int d^2q_{\perp}\frac{1}{[(\vec{q}_{\perp}-\vec{k}_{\perp})^2+m_D^2]}\delta^2(\vec{q}_{Tn}-\vec{q}_{\perp})
\eea
\item
Diagrams in Fig. \ref{Real} (c) and (d)  evaluate to 0. \\
\item
We now consider the diagram (e) including the mirror diagram
\small
\bea
  G_{e}^R&=& \frac{g^2N_c}{4}\nu^{\eta}\int \frac{d^2q_{\perp}dq^-}{(2\pi)^3q^-}\frac{Q|q^-|^{-\eta}}{Q(\frac{(\vec{q}_{\perp}+\vec{k}_{\perp})^2}{Q-q^-}+\frac{\vec{q}_{\perp}^2+m^2}{q^-})-\vec{k}_{\perp}^2}\frac{1}{q^-(-\frac{(\vec{q}_{\perp}+\vec{k}_{\perp})^2}{Q-q^-})-(\vec{q}_{\perp}+\vec{k}_{\perp})^2-m^2} \nn\\
&&\Bigg[\frac{-8(\vec{q}_{\perp}+\vec{k}_{\perp})\cdot \vec{q}_{\perp}}{q^-} - 4\frac{(\vec{q}_{\perp}+\vec{k}_{\perp})^2}{Q-q^-}-4\vec{k}_{\perp}\cdot \frac{(\vec{q}_{\perp}+\vec{k}_{\perp})}{Q}- 4q^-\frac{(\vec{q}_{\perp}+\vec{k}_{\perp})^2}{(Q-q^-)^2}-4\vec{q}_{\perp}\cdot \frac{(\vec{q}_{\perp}+\vec{k}_{\perp})}{Q-q^-}\Bigg]\mathcal{M}\nn\\
\eea
\normalsize
Only the first term in the square brackets contributes to the rapidity divergence 
\bea
 G_{e,\eta}^{R}&=&-2g^2N_c(\nu/Q)^{\eta}\int_0^{z_{c}} \frac{dz}{z^{1+\eta}(2\pi)^3}\int d^2q_{\perp}\frac{1}{(\vec{q}_{\perp}-\vec{k}_{\perp})^2+m^2}\frac{1}{(\vec{q}_{\perp})^2+m^2} \nn\\
&\times& \Bigg[(\vec{q}_{\perp})\cdot (\vec{q}_{\perp}-\vec{k}_{\perp})\Bigg]\delta^2(\vec{q}_{Tn}-\vec{q}_{\perp})\nn\\
&=& 2g^2N_c(\nu/[Qz_c])^{\eta}\frac{1}{\eta}\frac{1}{(2\pi)^3}\int d^2q_{\perp}\frac{\vec{q}_{\perp}\cdot (\vec{q}_{\perp}-\vec{k}_{\perp})}{[(\vec{q}_{\perp}-\vec{k}_{\perp})^2+m^2][(\vec{q}_{\perp})^2+m^2]}\delta^2(\vec{q}_{Tn}-\vec{q}_{\perp})\nn
\eea

\item
We can similarly compute the diagram Fig. \ref{Real}(f), which gives us  
\bea
  G_{f}^R&=& -\nu^{\eta}\frac{g^2}{4}N_c\int \frac{d^2q_{\perp}dq^-}{q^-(2\pi)^3}\frac{Q-q^-}{(Q-q^-)(-\frac{q_{\perp}^2+m^2}{q^-})-\vec{q}_{\perp}^2}\frac{|q^-|^{\eta}}{q^-(-\frac{(\vec{q}_{\perp}+\vec{k}_{\perp})^2}{Q-q^-})-(\vec{q}_{\perp}+\vec{k}_{\perp})^2-m^2} \nn\\
&\times& \Bigg[\frac{8(\vec{q}_{\perp}+\vec{k}_{\perp})\cdot \vec{q}_{\perp}}{q^-} + 8(\vec{q}_{\perp}+\vec{k}_{\perp})\cdot\frac{(\vec{q}_{\perp}}{Q-q^-}+4q^-\frac{(\vec{q}_{\perp}+\vec{k}_{\perp})}{(Q-q^-)^2}\cdot\vec{q}_{\perp}\Bigg]
\eea
The rapidity divergence only appears again only in the first term in the square brackets which gives us an identical result to diagram(e)
\bea
G_{f,2}^{R,\eta}&=&  2g^2N_c(\nu/[Qz_c])^{\eta}\frac{1}{\eta}\frac{1}{(2\pi)^3}\int d^2q_{\perp}\frac{\vec{q}_{\perp}\cdot (\vec{q}_{\perp}-\vec{k}_{\perp})}{[(\vec{q}_{\perp}-\vec{k}_{\perp})^2+m^2][(\vec{q}_{\perp})^2+m^2]}\delta^2(\vec{q}_{Tn}-\vec{q}_{\perp})\nn
\eea
\item
Next we consider the diagrams purely from Lagrangian insertions namely diagrams (g),(h) and (i), none of which have a rapidity divergence
\item
Finally we have the real diagram (j) which gives us
\bea
 G_{j}^R&=& -g^2\nu^{\eta}N_c\delta^{AB}\int dp^+ \delta( p^+-\frac{(\vec{q}_{\perp}+\vec{k}_{\perp})^2}{Q-q^-}) \int \frac{d^4q}{(2\pi)^3}\delta(q^2-m^2) \frac{p^-|q^-|^{-\eta}}{[(q+k)^2-m^2]^2}\nn\\
&\times&(\vec{q}_{\perp}+\vec{k}_{\perp})^2\left(-4\frac{q^-}{p^-}-4-2\left(\frac{q^-}{p^-}\right)^2\right)\mathcal{M}
\eea

Putting in the measurement, we see that this piece also has a rapidity divergence, which we can isolate here 
 \bea
  G_{j,\eta}^{R} &= &   -2g^2N_c\delta^{AB}\left(\frac{\nu}{Qz_c}\right)^{\eta}\frac{1}{\eta}\frac{1}{(2\pi)^3} \int d^2q_{\perp}\frac{1}{\vec{q}_{\perp}^2+m^2}\delta^2(\vec{q}_{Tn}-\vec{q}_{\perp})
\eea
\end{itemize}

We therefore see that the real diagrams do not have any UV divergence but do contribute to the rapidity divergence. 
The rapidity divergent pieces can be collected together to give 
\bea
G_R^{\eta}&=& -\frac{2N_cg^2}{(2\pi)^3}\left(\frac{\nu}{Qz_c}\right)^{\eta}\frac{1}{\eta}\int \frac{d^2q_{\perp}\delta^{AB}\delta^2(\vec{q}_{Tn}-\vec{q}_{\perp})}{k_{\perp}^2}\Bigg\{\frac{1}{(\vec{q}_{\perp}-\vec{k}_{\perp})^2}+\frac{1}{\vec{q}_{\perp}^2}-\frac{2\vec{q}_{\perp}\cdot (\vec{q}_{\perp}-\vec{k}_{\perp})}{q_{\perp}^2(\vec{q}_{\perp}-\vec{k}_{\perp})^2}\Bigg\}\nn\\
&=& -\frac{2N_cg^2}{(2\pi)^3}\left(\frac{\nu}{Qz_c}\right)^{\eta}\frac{1}{\eta}\int \frac{d^2q_{\perp}\delta^{AB}\delta^2(\vec{q}_{Tn}-\vec{q}_{\perp})}{[q_{\perp}^2+m_D^2][(\vec{q}_{\perp}-\vec{k}_{\perp})^2+m_D^2]}
\label{RealRapid}
\eea
The natural scale is $\nu \sim Q \sim Qz_{c}$.
 
\begin{figure}
  \includegraphics[width=\textwidth]{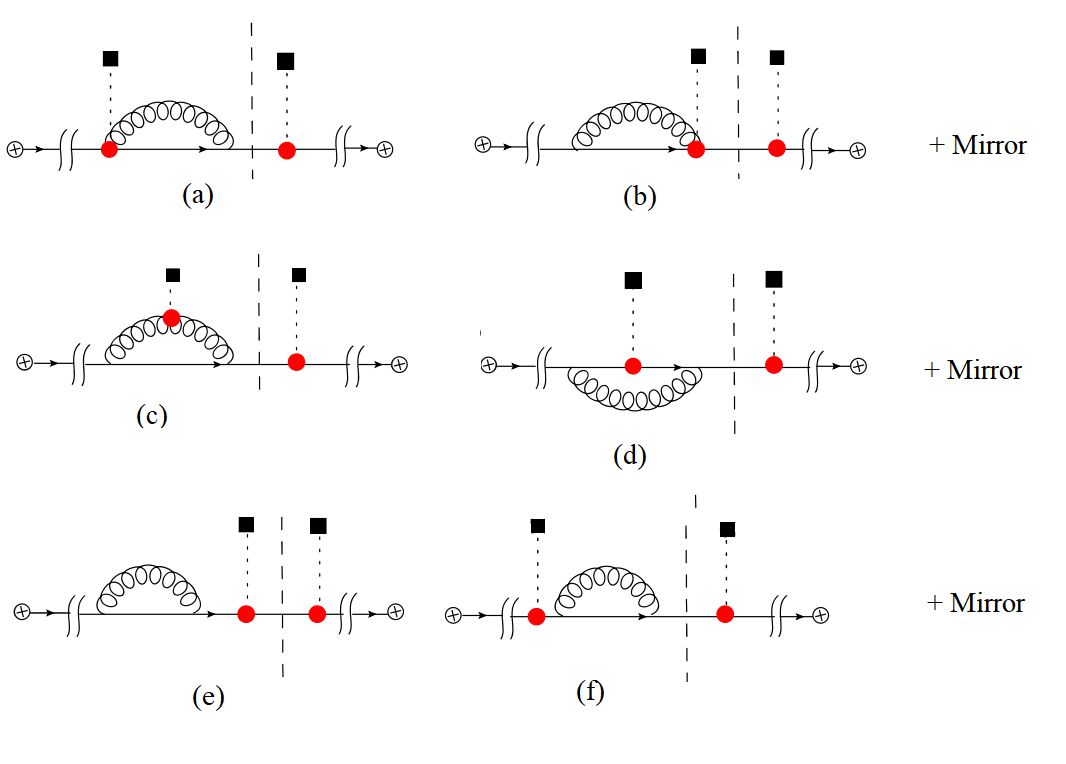}
  \caption{Virtual gluon diagrams for $\mathcal{J}^{AB}_{n,R}$.}
  \label{Virtual}
\end{figure}

We also have virtual diagrams which are shown in Fig.\ref{Virtual} but these have already been computed in \cite{Rothstein:2016bsq}, so we will take the result from there directly. 

The virtual diagrams again do not have any UV divergences but only rapidity divergences.  Keeping only the divergent terms, 
\bea
 V_R^{\eta}&=& -\frac{N_cg^2}{(2\pi)^3}\left(\frac{\nu}{Q}\right)^{\eta}\int_0^{1}\frac{dz}{z^{1+\eta}}\int \frac{d^2q_{\perp}\delta^{AB}\delta^2(\vec{q}_{Tn}-\vec{k}_{\perp})}{q_{\perp}^2(\vec{q}_{\perp}-\vec{k}_{\perp})^2}\nn\\
&=&  \frac{N_cg^2}{(2\pi)^3}\left(\frac{\nu}{Q}\right)^{\eta}\frac{1}{\eta}\int \frac{d^2q_{\perp}\delta^{AB}\delta^2(\vec{q}_{Tn}-\vec{k}_{\perp})}{q_{\perp}^2(\vec{q}_{\perp}-\vec{k}_{\perp})^2}
\label{VirtualRapid}
\eea


\subsubsection{$\mathcal{J}_{n,V}^{AB}$}
The set of diagrams to be computed for this piece remains the same, the only difference is the measurement function, hence we can immediately write down the rapidity divergent result 
\bea
\mathcal{J}_{n,V}^{AB}&=&  \delta^{AB}\delta^2(\vec{q}_{Tn})\frac{2N_cg^2}{(2\pi)^3}\left(\frac{\nu}{Q}\right)^{\eta}\int_0^{1}\frac{dz}{z^{1+\eta}}\left(\frac{1}{q_{\perp}^2(\vec{q}_{\perp}-\vec{k}_{\perp})^2}-\frac{1}{2q_{\perp}^2(\vec{q}_{\perp}-\vec{k}_{\perp})^2}\right)
\eea
Using this result along with Eqs. \ref{RealRapid} and \ref{VirtualRapid} in  Eq.\ref{JetMed}, we can now write 
\bea
 \nu\frac{d}{d\nu} \mathcal{J}_{n,\text{Med}}^{AB(1)}(\vec{k}_{\perp})= -\frac{\alpha_sN_c}{\pi^2}\int d^2q_{\perp} \left( \frac{\mathcal{J}_{n,\text{Med}}^{AB(0)}(\vec{q}_{\perp})}{(\vec{q}_{\perp}-\vec{k}_{\perp})^2} -\frac{k_{\perp}^2\mathcal{J}_{n,\text{Med}}^{AB(0)}(k_{\perp})}{2q_{\perp}^2(\vec{q}_{\perp}-\vec{k}_{\perp})^2}\right)
\eea
The result therefore gives us the final Rapidity RGE for the Jet function at this perturbative order
 \bea
\nu\frac{d}{d\nu} J(\vec{k}_{\perp})= -\frac{\alpha_sN_c}{\pi^2}\int d^2q_{\perp} \left( \frac{J(\vec{q}_{\perp})}{(\vec{q}_{\perp}-\vec{k}_{\perp})^2} -\frac{k_{\perp}^2J(k_{\perp})}{2q_{\perp}^2(\vec{q}_{\perp}-\vec{k}_{\perp})^2}\right)
\eea
This is identical to the BFKL equation with a negative sign. The medium induced jet function does not yield any UV anomalous dimension.


\subsection{Medium Soft function} 
The medium soft function defined in Eq.\ref{MedSoft} is just the Wightman function in a thermal bath. 
To compute the corrections upto NLO, I will use the realtime formalism of thermal field theory. Operationally, this involves introducing two copies of every field in our action dubbing them as type 1 and type 2 fields. The two types of fields only talk to each other via the kinetic term, which allows type 1 to propagate into type 2 and vice versa. All the fields in an interaction term have either a type 1 or type 2 field with opposite sign of the coupling for the interaction term with type 2 fields. In my case, for simplicity I assume that the thermal bath is only sourcing quarks so that I only double the quark fields. This is sufficient to derive the RG equations for the soft function. Including the gluons sourced by the thermal medium will only change the boundary condition for the solution of the RG equations. Doubling of the quark fields leads to 4 types of fermion propagators 
\bea
D_{11}(k) &=&\slashed{k}\Bigg[\frac{i}{k^2+i\epsilon}-2\pi\delta(k^2)n_F(|k^0|)\Bigg] , \ \ \ \ D_{22}(k)= \slashed{k}\Bigg[\frac{-i}{k^2-i\epsilon}-2\pi\delta(k^2)n_F(|k^0|)\Bigg]\nn\\
D_{21}(k) &=& 2\pi\slashed{k}\Bigg[\Theta(k^0)\delta(k^2)-\delta(k^2)n_F(|k^0|)\Bigg], \ \ \ \ D_{12}(k)= 2\pi\slashed{k}\Bigg[\Theta(-k^0)\delta(k^2)-\delta(k^2)n_F(|k^0|)\Bigg]\nn
\eea
where $n_F(\omega) = \frac{1}{e^{\beta \omega}+1}$ is the Fermi distribution function.

\begin{figure}
\centering
  \includegraphics[width=0.3\linewidth]{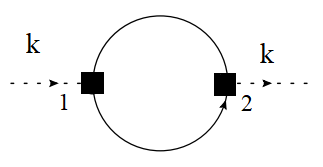}
  \caption{The tree level medium structure function in a thermal bath is just the advanced Wightman function. The black square is the Soft Glauber current vertex.}
  \label{Stree}
\end{figure}

The tree level diagram is shown in Fig. \ref{Stree} which gives us 

\bea
&&S_G^{AB,(0)}= -\frac{2(2\pi)^3}{k_{\perp}^2} \int \frac{dk^-}{(2\pi)}\int \frac{d^4p}{(2\pi)^3}(p^+)^2\Big[\delta^-(p^2)-\delta(p^2)n_F(|p^0|)\Big]\nn\\
&&\Big[\delta^+((p+k)^2)-\delta((p+k)^2)n_F(|p^0+k^0|)\Big]\text{Tr}\Big[ \frac{\slashed{n}}{2}\frac{\slashed{\bar n}}{2}\Big]\text{Tr}[T^AT^B]
\eea

which evaluates to 
\bea
S_G^{AB,(0)}(\vec{k}_{\perp})&=& \frac{\delta^{AB}}{k_{\perp}^2} \int d^2p_{\perp}\int_0^{\infty} dp^+ \Bigg\{n_F\left(|\frac{p^+}{2}+\frac{p_{\perp}^2}{2}|\right)\Big[1-n_F(|\frac{p^+}{2}+\frac{(\vec{p}_{\perp}+\vec{k}_{\perp})^2}{2p^+}|)\Big]\nn\\
&+&n_F(|\frac{p^+}{2}+\frac{(\vec{p}_{\perp}+\vec{k}_{\perp})^2}{2p^+}|)\Big[1-n_F\left(|\frac{p^+}{2}+\frac{p_{\perp}^2}{2}|\right)\Big]\Bigg\}\nn
\eea

We now turn to the one loop corrections. 	We can look at the real and virtual graphs separately, where by real graphs we mean those where the gluon propagator is of type $D_{12}$ or $D_{21}$. We will only compute the rapidity divergences to establish consistency of the Rapidity anomalous dimension. The UV divergence of the soft function can be obtained by consistency with the hard function since we have already established that the jet function does not contain any UV divergences.

\subsubsection{Real Soft Diagrams}
\begin{figure}
  \includegraphics[width=\linewidth]{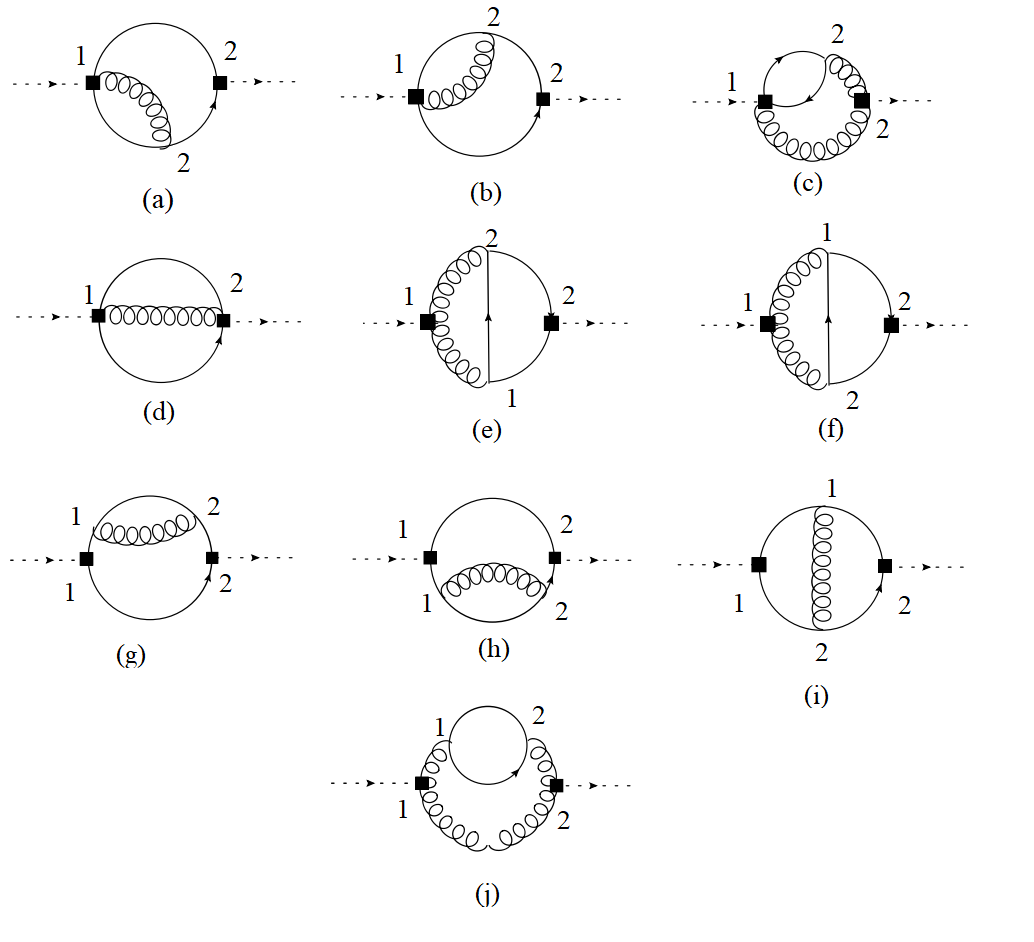}
  \caption{"Real" gluon emission diagrams for the medium structure function. The black square is the Soft Glauber vertex. The real diagrams are those where the gluon propagator is of type $D_{12}$ or $D_{21}$, i.e, the non-thermal piece of the propagator is on-shell. }
  \label{SoftR}
\end{figure}

 All the real graphs are shown in Fig. \ref{SoftR}. We also include the mirror diagrams not shown explicitly in this figure.
\begin{itemize}
\item Diagram (a) involves the soft Wilson line along with an insertion from the Soft Hamiltonian which is identical to the full QCD Hamiltonian.
\bea
S_a^R&= & -4N_cg^2\nu^{\eta/2}\frac{\delta^{AB}}{k_{\perp}^2}\int dk^-\int d^4q\int d^4p(p^+)^2\Big[\delta^-(p^2)-\delta(p^2)n_F(|p^0|)\Big]\delta^+(q^2-m^2)\nn\\
&\times&\Big[\delta^+((p+k-q)^2)-\delta((p+k-q)^2)n_F(|p^0+k^0-q^0|)\Big]\frac{p^+}{(p+k)^2}\frac{|q^--q^+|^{-\eta/2}}{q^+}\nn
\eea
p is the quark momentum whole q is the gluon momentum. k is the Glauber momentum exchanged with the jet. The overall minus sign comes from the Quark loop. In our case, there is no $\bar n$ sector for the medium interaction and hence the rapidity divergence only exists in the limit $q^- \rightarrow \infty$ when the soft sector overlaps with the n collinear sector. Hence we can simply use $|q^-|^{-\eta/2}$ as a rapidity regulator in place of the complete $|q^--q^+|^{\eta/2}$ which simplifies our computation. If we only keep track of the pieces with rapidity divergence, we end up with two terms 
\bea 
 S_{a,1}^{R,\eta}&=&g^2N_c\frac{\delta^{AB}}{k_{\perp}^2}\int d^2p_{\perp}\int_0^{\infty}dp^+ \int \frac{d^2q_{\perp}}{(q_{\perp}^2+m_D^2)^{\eta/2}}n_F(|\frac{p^+}{2}+\frac{p_{\perp}^2}{2p^+}|)\nn\\
&\times& (\nu p^+)^{\eta/2}\int_0^{1}\frac{dz}{(2\pi)^{d-1}}\frac{(1-z)}{z(\vec{p}_{\perp}+\vec{k}_{\perp}-\vec{q}_{\perp})^2+(1-z)[q_{\perp}^2+m_D^2]-z(1-z)(\vec{p}_{\perp}+\vec{k}_{\perp})^2}\frac{1}{(z)^{1-\eta/2}}\nn\\
&\times&\left(1-n_F(|\frac{p^+(1-z)}{2}+\frac{(\vec{p}_{\perp}+\vec{k}_{\perp}-\vec{q}_{\perp})^2}{2p^+(1-z)}|)\right)
\eea
and 
\bea
 S_{a,2}^{R,\eta}&=&g^2N_c\frac{\delta^{AB}}{k_{\perp}^2}\int d^2p_{\perp}\int_{0}^{\infty} dp^+\int \frac{d^2q_{\perp}}{(q_{\perp}^2+m_D^2)^{\eta/2}}\left(1-n_F(|\frac{p^+}{2}+\frac{p_{\perp}^2}{2p^+}|)\right)\nn\\
&\times& (\nu p^+)^{\eta/2}\frac{\int_0^{\infty}dz}{(2\pi)^{d-1}}\frac{(1-z)}{z(\vec{p}_{\perp}+\vec{k}_{\perp}-\vec{q}_{\perp})^2+(1-z)[q_{\perp}^2+m_D^2]-z(1-z)(\vec{p}_{\perp}+\vec{k}_{\perp})^2}\frac{1}{(z)^{1-\eta/2}}\nn\\
&\times&n_F(|\frac{p^+(1+z)}{2}+\frac{(\vec{p}_{\perp}+\vec{k}_{\perp}-\vec{q}_{\perp})^2}{2p^+(1+z)}|)
\eea

Due to the presence of the Fermi distribution function, it is not possible to do the integral over z analytically. However, if we are only interested in the rapidity pole, we see that this happens in the limit $z \rightarrow 0$, which allows us to isolate the rapidity pole
\bea
S_a^{R,\eta}&=& g^2N_c\frac{\delta^{AB}}{k_{\perp}^2}\frac{2}{\eta}\int d^2p_{\perp}dp^+ \int \frac{d^2q_{\perp}}{(2\pi)^{3}}\frac{1}{q_{\perp}^2+m_D^2}\left(\frac{\nu p^+}{q_{\perp}^2+m_D^2}\right)^{\eta/2}\nn\\
&&\Bigg\{n_F(|\frac{p^+}{2}+\frac{p_{\perp}^2}{2p^+}|)\left(1-n_F(|\frac{p^+}{2}+\frac{(\vec{p}_{\perp}+\vec{k}_{\perp}-\vec{q}_{\perp})^2}{2p^+}|)\right)\nn\\
 &+&(1-n_F(|\frac{p^+}{2}+\frac{p_{\perp}^2}{2p^+}|))n_F(|\frac{p^+}{2}+\frac{(\vec{p}_{\perp}+\vec{k}_{\perp}-\vec{q}_{\perp})^2}{2p^+}|)\Bigg\}
\eea

\item
We next consider the contribution from Diagram (b) 
\bea
S_b^R&= & 4g^2N_c\nu^{\eta/2}\frac{\delta^{AB}}{k_{\perp}^2}\int d^4p(p^+)^2\left(\delta^+((p+k-q)^2)-\delta((p+k-q)^2)n_F(|p^0+k^0-q^0|)\right)\nn\\
&\times&\left( \delta^-(p^2)-\delta(p^2)n_F(|p^0|)\right) \int d^4q \delta^+(q^2-m_D^2)\frac{p^+-q^+}{(p-q)^2}\frac{|q^-|^{-\eta/2}}{q^+}
\eea
The rapidity divergence contribution is identical to case (a).

\item{}
Diagrams (c) and (d) reduce to 0. 

\item{}
For the diagram (e), we have 
\small
\bea
  &&S_e^R= g^2N_c\nu^{\eta/2}\frac{\delta^{AB}}{k_{\perp}^2}\int d^4p\left( \delta^-(p^2)-\delta(p^2)n_F(|p^0|)\right)\left(\delta^+((p+k-q)^2)-\delta((p+k-q)^2)n_F(|p^0+k^0-q^0|)\right)\nn\\
&\times& \int d^4q \frac{\delta^+(q^2-m_D^2)p^+}{(p+k)^2[(q-k)^2-m^2]}\Big[\frac{8p^+(p^+-q^+)q_{\perp}\cdot(q_{\perp}-k_{\perp})|q^-|^{-\eta/2}}{q^+}\nn\\
&-&4p^+(2q_{\perp}-k_{\perp})\cdot(p_{\perp}+k_{\perp}-q_{\perp})-4(p^+-q^+)((p_{\perp}+k_{\perp})\cdot(q_{\perp}-k_{\perp})+p_{\perp}\cdot q_{\perp})\nn\\
&+&4p^+(2p_{\perp}+k_{\perp})\cdot(p_{\perp}+k_{\perp}-q_{\perp})-4p^+(p+k-q)\cdot(2p+k)+4(p^+-q^+)p\cdot(p+k)\Big]
\eea
\normalsize
 Only the first term in the square brackets above has a rapidity divergence. This yields a rapidity pole
 \bea
S_e^{R,\eta}&=&   2N_cg^2\frac{\delta^{AB}}{k_{\perp}^2}\int d^2p_{\perp}dp^+\int \frac{d^2q_{\perp}}{(2\pi)^3}\frac{q_{\perp}\cdot(q_{\perp}-k_{\perp})}{(q_{\perp}^2+m_D^2)((\vec{q}_{\perp}-\vec{k}_{\perp})^2+m_D^2)}\nn\\
&&\left(\frac{\nu p^+}{q_{\perp}^2+m_D^2}\right)^{\eta/2}\frac{2}{\eta}\Bigg\{n_F(|\frac{p^+}{2}+\frac{p_{\perp}^2}{2p^+}|)\left(1-n_F(|\frac{p^+}{2}+\frac{(\vec{p}_{\perp}+\vec{k}_{\perp}-\vec{q}_{\perp})^2}{2p^+}|)\right)\nn\\
 &+&(1-n_F(|\frac{p^+}{2}+\frac{p_{\perp}^2}{2p^+}|))n_F(|\frac{p^+}{2}+\frac{(\vec{p}_{\perp}+\vec{k}_{\perp}-\vec{q}_{\perp})^2}{2p^+}|)\Bigg\}
\eea

\item
Diagram(f) gives the same contribution as diagram (e)
\item
Diagram (g) only has insertions from the Soft Hamiltonian and evaluates to 
\bea
S_g^{R}&=& g^2C_F\delta^{AB}\int d^2p_{\perp}\frac{dp^+}{p^+} \int \frac{d^4q}{(2\pi)^3} \delta^+(q^2-m^2)\frac{\Big\{-2(p-q)^2p^+ +4p\cdot(p-q)(p^+-q^+)\Big\}}{[(p-q)^2]^2}\nn\\
&\times& \left( \Theta(-p^+)-\text{sgn}(-p^+)n_F(|p^0|)\right)\left(\Theta(p^+-q^+)-\text{sgn}(p^+-q^+)n_F(|p^0+k^0-q^0|)\right)
\eea
This piece does not have a rapidity divergence and hence we will not compute it further.

\item
\bea
S_{h}^R &=&-C_F\frac{g^2}{4}\int d^2p_{\perp}dp^+\int \frac{d^2q_{\perp}dq^+}{(2\pi)^3q^+(p^+-q^+)}\frac{\left(4(p+k)^2(p^+-q^+)-8p^+(p+k)\cdot(p+k-q)\right)}{[(p+k)^2]^2}\nn\\
&\times& \left( \Theta(-p^+)-\text{sgn}(-p^+)n_F(|p^0|)\right)\left(\Theta(p^+-q^+)-\text{sgn}(p^+-q^+)n_F(|p^0+k^0-q^0|)\right)
\eea
which is also rapidity finite.

\item
Same is the case for (i) which has the form
\small
\bea
 S_i^R&=&-2g^2\delta^{AB}\left(\frac{C_F}{2}-\frac{Nc}{4}\right)\int d^2p_{\perp}\frac{dp^+}{p^+}\int \frac{d^4q\delta^+(q^2-m^2)}{(p^+-q^+)(2\pi)^3}\frac{\left(-2(p+k)^2q^+(p^+-q^+)+2(p-q)^2p^+q^+\right)}{[(p+k)^2][(p-q)^2]}\nn\\
&&\left( \Theta(-p^+)-\text{sgn}(-p^+)n_F(|p^0|)\right)\left(\Theta(p^+-q^+)-\text{sgn}(p^+-q^+)n_F(|p^0+k^0-q^0|)\right)
\eea
\normalsize
\item
We finally have (j) which does have a rapidity divergence
\small
\bea
 &&S_j^R= -g^2\nu^{\eta/2}\frac{N_c}{4}\frac{\delta^{AB}}{k_{\perp}^2}\int d^2p_{\perp}\frac{dp^+}{p^+}\int \frac{d^2q_{\perp}dq^+}{(2\pi)^3(p^+-q^+)} \frac{|\frac{q_{\perp}^2+m_D^2}{q^+}|^{-\eta/2}}{q^+[q^+(\frac{p_{\perp}^2}{p^+}-\frac{(\vec{p}_{\perp}+\vec{k}_{\perp}-\vec{q}_{\perp})^2}{p^+-q^+}-(\vec{q}_{\perp}-\vec{k}_{\perp})^2-m^2]^2}\nn\\
&&\left(\Theta(-p^+)-\text{sgn}{-p^+}n_F(|p^0|)\right)\left(\Theta(p^+-q^+)-\text{sgn}(p^+-q^+)n_F(|p^0+k^0-q^0|)\right)\nn\\
&&\Big\{-8q^+p^+(q_{\perp}-k_{\perp})\cdot(p_{\perp}+k_{\perp}-q_{\perp})+8p^+(p^+-q^+)(q_{\perp}-k_{\perp})\cdot(q_{\perp}-k_{\perp})\nn\\
&+&8(q^+)^2(p_{\perp}\cdot(p_{\perp}+k_{\perp}-q_{\perp})-p\cdot(p+k-q))-8q^+(p^+-q^+)p_{\perp}\cdot(q_{\perp}-k_{\perp})\Big\}
\eea
\normalsize
As before, we can isolate the rapidity divergence 
\bea
 S_j^{R,\eta}&=&   2g^2N_c\frac{\delta^{AB}}{k_{\perp}^2}\Bigg\{\int d^2p_{\perp}dp^+\int \frac{d^2q_{\perp}}{(2\pi)^{d-1}}n_F(|\frac{p^+}{2}+\frac{p_{\perp}^2}{2p^+}|)\left(1-n_F(|\frac{p^+}{2}+\frac{(\vec{p}_{\perp}+\vec{k}_{\perp}-\vec{q}_{\perp})^2}{2p^+}|)\right)\nn\\
 &+&\int d^2p_{\perp}dp^+(1-n_F(|\frac{p^+}{2}+\frac{p_{\perp}^2}{2p^+}|))n_F(|\frac{p^+}{2}+\frac{(\vec{p}_{\perp}+\vec{k}_{\perp}-\vec{q}_{\perp})^2}{2p^+}|)\Bigg\}\nn\\
&\times&\left(\frac{\nu p^+}{q_{\perp}^2+m_D^2}\right)^{\eta/2}\frac{2}{\eta}\frac{1}{(\vec{q}_{\perp}-\vec{k}_{\perp})^2+m_D^2}
\eea

\end{itemize}

We can combine all the rapidity pole pieces above and make the shift $\vec{q}_{\perp} \rightarrow -\vec{q}_{\perp}+\vec{k}_{\perp}$ to write
\small
\bea
\label{RapidSoftR}
&&S^{R,\eta} = \delta^{AB}\frac{\alpha_sN_c}{\pi^2k_{\perp}^2}\int d^2p_{\perp}dp^+ \int d^2q_{\perp} \left(\frac{\nu p^+}{(\vec{q}_{\perp}-\vec{k}_{\perp})^2+m_D^2}\right)^{\eta/2}\frac{2}{\eta}\frac{k_{\perp}^2}{q_{\perp}^2 (\vec{q}_{\perp}-\vec{k}_{\perp})^2}\nn\\
&&\Bigg\{ n_F(|\frac{p^+}{2}+\frac{p_{\perp}^2}{2p^+}|)\Big[1- n_F\left( \frac{p^+}{2}+\frac{(\vec{p}_{\perp}+\vec{q}_{\perp})^2}{2p^+}|\right)\Big]+ n_F(|\frac{p^+}{2}+\frac{(\vec{p}_{\perp}+\vec{q}_{\perp})^2}{2p^+}|)\Big[1- n_F\left( \frac{p^+}{2}+\frac{\vec{p}_{\perp}^2}{2p^+}|\right)\Big] \Bigg\}\nn\\
\eea
\normalsize 
so that the result is now written in terms of the tree level soft function.

\subsubsection{Virtual Soft diagrams}

\begin{figure}
  \includegraphics[width=\linewidth]{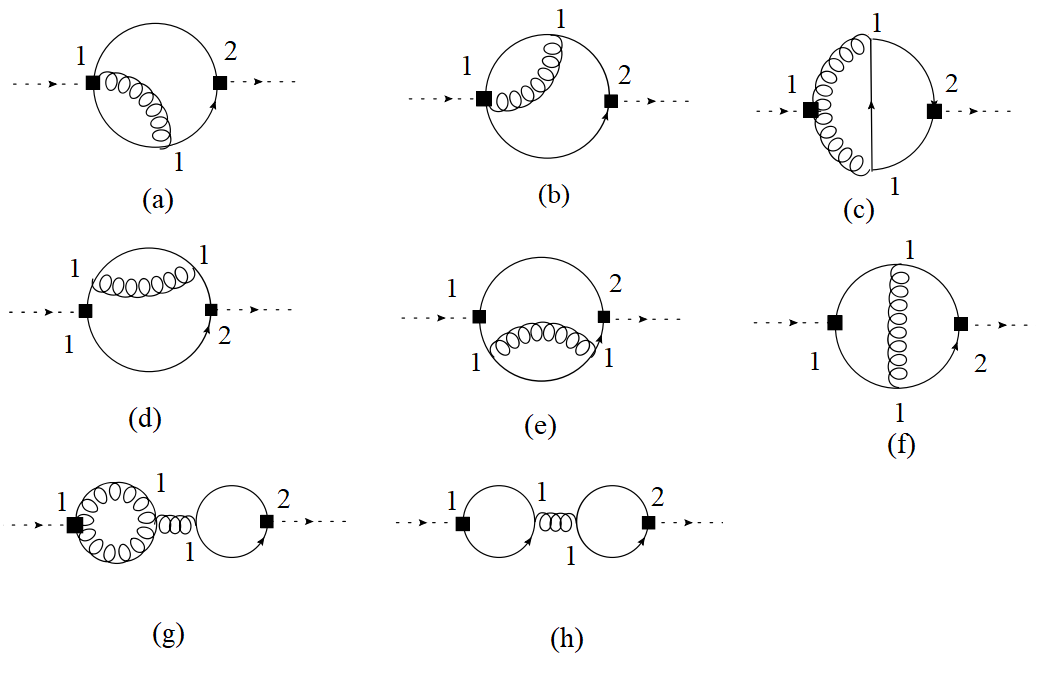}
  \caption{"Virtual" gluon emission for medium structure function. The gluon propagator is of type $D_{11}$ or $D_{22}$ meaning the non-thermal piece of the propagator is just the zero temperature Feynman propagator.}
  \label{SoftV}
\end{figure}

The virtual diagrams are the one where the gluon propagator is of type $D_{11}$, $D_{22}$. All the possible diagrams in this category are shown in Fig. \ref{SoftV}

As for the real diagrams, we concentrate on isolating the rapidity divergences in each diagram.

\begin{itemize}
\item
Diagram (a) evaluates to 
\bea
&&S_a^V=\frac{N_c}{4}\nu^{\eta/2}16g^2\int d^4p p^+(\delta^-(p^2)-\delta(p^2)n_F(|p^0|))\int \frac{d^dq}{(2\pi)^{d-1}}\frac{|q^-|^{-\eta/2}}{q^+}p^+\Big\{\frac{i(p^+-q^+)}{(p+k-q)^2} \nn\\
&&-(p^+-q^+)\delta((p+k-q)^2)n_F(|p^0+k^0-q^0)\Big\}\Big\{\delta^+((p+k)^2)-\delta((p+k)^2)n_F(|p^0+k^0|)\Big\}\nn
\eea
This is now proportional to the tree level result and gives us a rapidity divergence 
\bea
 S_a^{V,\eta}&=&g^2N_c\int d^2p_{\perp}\int_0^{\infty}dp^+\Bigg\{ n_F(|p^0|))\Big\{1-n_F(|p^0+k^0|)\Big\}+ (1-n_F(|p^0|))n_F(|p^0+k^0|)\Bigg\}\nn\\
&\times&\int \frac{d^2q_{\perp}}{(2\pi)^{d-1}}\frac{1}{q_{\perp}^2+m_D^2}\left(\frac{\nu p^+}{q_{\perp}^2+m_D^2}\right)^{\eta/2}\frac{2}{\eta}
\eea
where $p^0 = \frac{p^+}{2}+ \frac{p_{\perp}^2}{2p^+}$ and $ p^0+k^0 = \frac{p^+}{2}+\frac{(\vec{p}_{\perp}+\vec{k}_{\perp})^2}{2p^+}$ enforcing on-shell conditions.
\item
Diagram (b) evaluates to 
\bea
S_b^V&=&\frac{N_c}{4}\nu^{\eta/2}16g^2\int d^4p p^+(\delta^-(p^2)-\delta(p^2)n_F(|p^0|))\int \frac{d^dq}{(2\pi)^{d-1}}\frac{i(p^++q^+)}{(p+q)^2}\frac{1}{q^2-m_D^2+i\epsilon}\nn\\
&\times& \frac{|q^-|^{-\eta/2}}{q^+}p^+\Big\{\delta^+((p+k)^2)-\delta((p+k)^2)n_F(|p^0+k^0|)\Big\}
\eea
Evaluating this tells us that the rapidity divergence is identical to that of diagram (a). 
\item 
Diagram (c) has the same trace structure as the corresponding real diagram and hence we can use that result directly 
\small 
\bea
  &&S_c^V= g^2N_c\nu^{\eta/2}\frac{\delta^{AB}}{k_{\perp}^2}\int d^4p\left( \delta^-(p^2)-\delta(p^2)n_F(|p^0|)\right)\left(\delta^+((p+k)^2)-\delta((p+k)^2)n_F(|p^0+k^0|)\right)\nn\\
&\times& \int d^4q \frac{|q^-|^{-\eta/2}}{[q^2-m^2]\left((p^+-q^+)(\frac{(\vec{p}_{\perp}+\vec{k}_{\perp})^2}{p^+}-q^-)-(\vec{p}_{\perp}+\vec{k}_{\perp}-\vec{q}_{\perp})^2\right)}\nn\\
&&\frac{1}{[q^+(q^--\frac{(\vec{p}_{\perp}+\vec{k}_{\perp})^2}{p^+}+\frac{p_{\perp}^2}{p^+})-(\vec{q}_{\perp}-\vec{k}_{\perp})^2 -m_D^2]}\Big[\frac{8p^+(p^+-q^+)q_{\perp}\cdot(q_{\perp}-k_{\perp})}{q^+}\nn\\
&&-4p^+(2q_{\perp}-k_{\perp})\cdot(p_{\perp}+k_{\perp}-q_{\perp})-4(p^+-q^+)((p_{\perp}+k_{\perp})\cdot(q_{\perp}-k_{\perp})+p_{\perp}\cdot q_{\perp})\nn\\
&+&4p^+(2p_{\perp}+k_{\perp})\cdot(p_{\perp}+k_{\perp}-q_{\perp})-4p^+(p+k-q)\cdot(2p+k)+4(p^+-q^+)p\cdot(p+k)\Big]
\eea
\normalsize
We can now isolate the rapidity divergence 
\bea
 S_c^{V,\eta}&=& -2g^2N_c\frac{\delta^{AB}}{k_{\perp}^2}\int d^2p_{\perp}\int_0^{\infty}dp^+\Bigg\{ n_F(|p^0|))\Big\{1-n_F(|p^0+k^0|)\Big\}+ (1-n_F(|p^0|))n_F(|p^0+k^0|)\Bigg\}\nn\\
&\times&\int\frac{ d^2q_{\perp}}{(2\pi)^3}\frac{q_{\perp}\cdot(q_{\perp}-k_{\perp})}{[(\vec{q}_{\perp}-\vec{k}_{\perp})^2+m_D^2][q_{\perp}^2+m_D^2]}\left(\frac{\nu p^+}{q_{\perp}^2+m_D^2}\right)^{\eta/2}\frac{2}{\eta}
\eea

\item
Diagrams (d) and (e) are simply wave-function renormalization for the quark and hence do not have any rapidity divergence. 
\item 
Diagram (f) evaluates to 
\small
\bea
 S_{f}^V &=& 4g^2 \text{Tr}[T^CT^AT^CT^B] \int d^2p_{\perp}dp^+\left( \Theta(-p^+)-\text{sgn}(-p^+)n_F(|p^0|)\right)\left(\Theta(p^+)-\text{sgn}(p^+)n_F(|p^0+k^0|)\right)\nn\\
&\times&  \int\frac{ d^4q}{(p^+)^2} \frac{1}{[q^2-m^2][(p+k-q)^2][(p-q)^2]}4\Big[(p^+-q^+)^2p\cdot(p+k)\nn\\
&+&(p^+)^2(p-q)\cdot(p+k-q)-p^+(p^+-q^+)[(p+k)\cdot(p+k-q)+p\cdot(p-q)]\Big]\nn
\eea
\normalsize
which gives a UV divergence but does not lead to a rapidity divergence. 
\item We also have wavefunction renormalization for the Glauber gluon which is shown in fig (g), (h)  which will contribute to only the UV divergence.
\end{itemize}
We can combine the Rapidity divergent pieces here to give 
\bea
\label{RapidSoftV}
S_V^{\eta}&=&g^2N_c\frac{\delta^{AB}}{k_{\perp}^2}\int d^2p_{\perp}\int_0^{\infty}dp^+\Bigg\{ n_F(|p^0|))\Big\{1-n_F(|p^0+k^0|)\Big\}+ (1-n_F(|p^0|))n_F(|p^0+k^0|)\Bigg\}\nn\\
&\times& (\nu p^+)^{\eta/2}\int\frac{d^2q_{\perp}}{(2\pi)^3}\left(\frac{-2q_{\perp}\cdot(q_{\perp}-k_{\perp})}{[q_{\perp}^2+m_D^2]^{1+\eta/2}[(\vec{q}_{\perp}-\vec{k}_{\perp})^2+m_D^2]}+\frac{2}{[q_{\perp}^2+m_D^2]^{1+\eta/2}}\right)\frac{2}{\eta}
\eea

Since we have neglected any contribution except the one with the rapidity pole, it is not obvious what is the natural scale for $\nu$. This will be, in general hard to answer by a direct calculation due to the presence of the Fermi distributions function. However, we know that the soft function does not know about the scale of the transverse momentum measurement $\vec{q}_T$, but only depends on $k_{\perp}$ and on T due to the presence of the density matrix. Hence, we conjecture that the natural scale for the soft function will be $\vec{k}_{\perp} \sim T$. However, we have an additional scale $m_D$ which is hierarchically separated from T and the tree level Soft function does provide support for $k_{\perp}$ all the way down to $m_D$. Hence a more rigorous treatment requires a further factorization of the scale T from the scale $m_D$ which can be done by matching the current EFT to an EFT that lives at $k_{\perp} \sim m_D$. This is beyond the scope of the current paper, but will be taken up in the future. For now, for the purposes of numerical analysis, we will assume that the scale for the Soft function is $k_{\perp}$. For the phenomenologically interesting case of $m_D \sim T$, there is a single scaling for the soft function but in that case, we expect the rapidity evolution to have a different form, where we incorporate the exact form of the mass dependent gluon propagator instead of simply using $m_D$ as an IR regulator.

\subsubsection{Soft function anomalous dimension}
We can now combine the results of the real and virtual diagrams( Eqs. \ref{RapidSoftR} and \ref{RapidSoftV}) to write the rapidity anomalous dimension for the Soft function at this order in perturbation theory.
 \bea
\nu\frac{d}{d\nu} S(\vec{k}_{\perp})= \frac{\alpha_sN_c}{\pi^2}\int d^2q_{\perp} \left( \frac{S(\vec{q}_{\perp})}{(\vec{q}_{\perp}-\vec{k}_{\perp})^2} -\frac{k_{\perp}^2S(k_{\perp})}{2q_{\perp}^2(\vec{q}_{\perp}-\vec{k}_{\perp})^2}\right)
\eea
This is just the BFKL equation and the resulting rapidity anomalous dimension is equal and opposite to the medium jet function so that we have a powerful check on the consistency of soft-collinear factorization.
From explicit calculation we already know that the jet function has no UV divergence and hence we can infer the UV anomalous dimension of the Soft function by consistency of factorization knowing that the hard function is just $\alpha^2(\mu)$.
\bea
 \mu\frac{d}{d\mu} S(\vec{k}_{\perp})&=& \frac{\alpha_s\beta_0}{\pi}S(k_{\perp})
\eea
where $\beta_0 = 11/3 C_A - C_Fn_FT_F$, which has a value of 23/3 if we assume 5 active quark flavors.

\section{Resummation}
\label{sec:Resum}
\begin{figure}
\centering
  \includegraphics[width=0.5\linewidth]{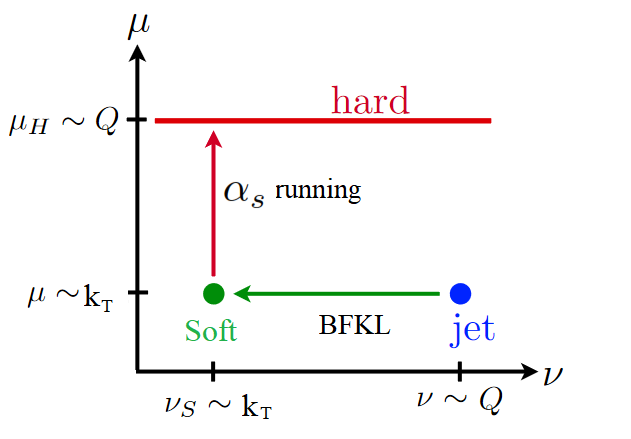}
  \caption{Path in $\mu$, $\nu$ space for Renormalization Group evolution of the medium structure function S and the medium induced jet function J.}
  \label{RGE}
\end{figure}

We can now solve the renormalization group equations in both $\mu, \nu$. The factorization of Hard-Soft-Collinear modes has completely separated out the physics at the scale Q from the IR scales $q_T$, $T$ and $k_{\perp}$. The solutions of the RG equations allow us to resum large logarithms in $Q/k_{\perp}$ thereby increasing the precision of the prediction. We run the jet function in $\nu$ and the Soft function in $\mu$ as shown in Fig. \ref{RGE}.

The rapidity RGE is the same as the BFKL equation and we first turn towards solving this equation.

\subsection{Solving the BFKL equation}
We follow the procedure outlined in \cite{Kovchegov:2012mbw}. We evolve the jet function using the BFKL equation from its natural scale $\nu \sim Q \sim Qz_c$ to the natural rapidity scale of the soft function $\nu \sim k_{\perp}$.  
\bea
\nu\frac{d}{d\nu} J(\vec{k}_{\perp})= -\frac{\alpha_sN_c}{\pi^2}\int d^2q_{\perp} \left( \frac{J(\vec{q}_{\perp})}{(\vec{q}_{\perp}-\vec{k}_{\perp})^2} -\frac{k_{\perp}^2J(k_{\perp})}{2q_{\perp}^2(\vec{q}_{\perp}-\vec{k}_{\perp})^2}\right)
\eea

We define the BFKL kernel as 
\bea
\int d^2q_{\perp}K_{BFKL}(q,k) J(q_{\perp}) = \frac{1}{\pi} \int \frac{d^2q_{\perp}}{(\vec{k}_{\perp}-\vec{q}_{\perp})^2}\Bigg[J(q_{\perp})-\frac{k_{\perp}^2}{2q_{\perp}^2}J(k_{\perp})\Bigg] 
\eea

So $K_{BFKL}$ defines for us the matrix in $q_{\perp}$ space which needs to be diagonalized. To do that, we need to find the eigenvalues and eigenfunctions of this matrix. The BFKL kernel is conformally invariant and its Eigenfunctions are $f(\vec{k}_{\perp})=k_{\perp}^{2\gamma-1}e^{i n\phi_k}$ with n being an integer and $\phi_k$ is the azimuthal angle for $\vec{k}_{\perp}$. If we plug in this ansatz in the equation above and do some manipulations, 
\bea
 \int d^2q_{\perp}K_{BFKL}(k,q)q_{\perp}^{2(\gamma-1)}e^{in\phi_q}= \chi(n,\gamma)k_{\perp}^{2(\gamma-1)}e^{i n\phi_k}
\eea

where 
\bea
\chi(n,\gamma) = \int_0^{\infty} dt\Bigg[\frac{1}{2\pi}\int_0^{2\pi} \frac{d\phi t^{\gamma-1}e^{in\phi}}{1+t-2\sqrt{t}\cos(\phi)}-\frac{1}{t}\left(\frac{1}{|t-1|}-\frac{1}{\sqrt{4t^2+1}}\right)\Bigg]
\eea
is the eigenvalue for the BFKL kernel which can be written as 
\bea
\chi(n,\gamma)=2 \psi(1)-\psi\left(\gamma+\frac{|n|}{2}\right)-\psi\left(1-\gamma+\frac{|n|}{2}\right)
\eea
where $\psi$ is the derivative of the logarithm of the $\Gamma$ function, also known as the Polygamma function. 
\bea
\psi(z) =\frac{d}{dz} \ln \Gamma(z) 
\eea
The result for the eigenvalue is valid for $0< Re(\gamma)<1$. \\
Given this result, we can now expand out our jet function in terms of these Eigenfunctions 
\bea
J(k_{\perp},\nu)=\sum_{n=-\infty}^{\infty} \int_{a-\infty}^{a+i\infty}\frac{d\gamma}{2\pi i}C_{n,\gamma}(\nu) k_{\perp}^{2(\gamma-1)}e^{in\phi_k}
\eea
We now plug this in our Rapidity Renormalization group equation 
\bea
\nu \frac{d}{d\nu}J(k_{\perp},\nu) &=& \sum_{n=-\infty}^{\infty}\int_{a-i\infty}^{a+i\infty}\frac{d\gamma}{2\pi i} k_{\perp}^{2(\gamma-1)}e^{in\phi} \Bigg[\nu \frac{d}{d\nu}C_{n,\gamma}(\nu)\Bigg] \nn\\
&=& -\frac{\alpha_sN_c}{\pi}\sum_{n=-\infty}^{\infty}\int_{a-i\infty}^{a+i\infty}\frac{d\gamma}{2\pi i} \chi(n,\gamma)C_{n,\gamma}(\nu)k_{\perp}^{2(\gamma-1)}e^{in\phi}\nn
\eea
Given that we have an expansion in a complete set of basis functions, we can write 
\bea
 \nu \frac{d}{d\nu}C_{n,\gamma}(\mu, \nu) = -\frac{\alpha_s(\mu) N_c}{\pi}\chi(n,\gamma)C_{n,\gamma}(\nu)
\eea
so that 
\bea
 C_{n,\gamma}(\nu_f) = C_{n,\gamma}(\mu, \nu_0)e^{ -\frac{\alpha_s(\mu) N_c}{\pi}\chi_{n,\gamma}\ln \frac{\nu_f}{\nu_0}} 
\eea
For the jet function, $\nu_0 \sim Q \sim Qz_{cut}$, so that our final solution looks like 
\bea
J(\mu, \nu_f, k_{\perp}) =\sum_{n=-\infty}^{\infty}\int_{a-i\infty}^{a+i\infty}\frac{d\gamma}{2\pi i} k_{\perp}^{2(\gamma-1)}e^{in\phi}  C_{n,\gamma}(\mu, Q)e^{ -\frac{\alpha_s(\mu) N_c}{\pi}\chi_{n,\gamma}\ln \frac{\nu_f}{Q}}  
\eea

We also know that at $\nu_f= Q$, the result reduces to the tree level result for the jet function which in transverse momentum space looks like 
\bea
J^{(0)}(\vec{k}_{\perp}, \vec{q}_{Tn}) = \frac{\delta^2(\vec{q}_{Tn}-\vec{k}_{\perp})}{k_{\perp}^2}-\frac{\delta(\vec{q}_{Tn})}{k_{\perp}^2} 
\eea
We therefore have the boundary condition 
\bea
 \frac{\delta^2(\vec{q}_{Tn}-\vec{k}_{\perp})}{k_{\perp}^2}-\frac{\delta^2(\vec{q}_{Tn})}{k_{\perp}^2} =\sum_{n=-\infty}^{\infty}\int_{a-i\infty}^{a+i\infty}\frac{d\gamma}{2\pi i} k_{\perp}^{2(\gamma-1)}e^{in\phi}  C_{n,\gamma}(\mu, Q)
\eea
which needs to be solved to fix $C_{n,\gamma}(Q)$. \\
We can first multiply both sides by $e^{-im\phi_k} k_{\perp}^{2\alpha^*-1}$ which is one element of the expansion basis and integrate over $\int d^2\vec{k}_{\perp}$.
We now us the orthogonality of the basis functions 
\bea
  \int d^2\vec{k}_{\perp}e^{-im\phi_k} k_{\perp}^{2(\alpha^*-1)}k_{\perp}^{2(\gamma-1)}e^{in\phi_k}= 2\pi \delta_{m,n}\int dr e^{(2\alpha_R+2\gamma_R-2)r}e^{i(-2\alpha_I+2\gamma_I)r}
\eea
To enforce orthogonality, we need to choose a single value for the real parts of $\alpha$ and $\gamma$ which now tells us that $\alpha_R=  \gamma_R =1/2$ which then gives us 
\bea
 \int d^2\vec{k}_{\perp}e^{-im\phi_k} k_{\perp}^{2(\alpha^*-1)}k_{\perp}^{2(\gamma-1)}e^{in\phi_k}=  2\pi^2 \delta_{m,n} \delta(\alpha_I-\gamma_I)
\eea
We therefore have 
\bea
\pi C_{m,\alpha}&=& e^{-im\phi_q}\frac{q_{tn}^{2(\alpha^*-1)}}{q_{Tn}^2}- 2\pi \delta^2(\vec{q}_{Tn})\delta_{m,0}\int dk_{\perp} \frac{k_{\perp}^{2(\alpha^*-1)}}{k_{\perp}}
\eea
The second term here essentially regulates the singularity as $k_{\perp} \rightarrow 0$. We can now write our result for our RG evolved jet function as 
\bea
J(\mu, \nu_f, k_{\perp}) &=&\sum_{n=-\infty}^{\infty}\int_{1/2-i\infty}^{1/2+i\infty}\frac{d\gamma}{2\pi^2 i} k_{\perp}^{2(\gamma-1)}e^{in\phi_k}\nn\\
&\times&\Bigg[e^{-in\phi_q}\frac{q_{Tn}^{2(\gamma^*-1)}}{q_{Tn}^2}- 2\pi\delta^2(\vec{q}_{Tn}) \delta_{n,0}\int dl_{\perp} \frac{l_{\perp}^{2(\alpha^*-1)}}{l_{\perp}}\Bigg] e^{ -\frac{\alpha_s(\mu) N_c}{\pi}\chi_{n,\gamma}\ln \frac{\nu_f}{Q}}
\label{JResum} 
\eea
\subsection{ Running the soft function} 
We can run the soft function in $\mu$ using the beta function 
\bea
S(k_{\perp},\nu \sim k_{\perp}, \mu \sim Q) &=& S(k_{\perp},\nu = k_{\perp}, \mu = k_{\perp}) e^{ \int_{k_{\perp}}^{Q} d \ln \mu \frac{\alpha_s(\mu)\beta_0}{\pi}} \nn\\
&=&  S(k_{\perp},\nu= k_{\perp}, \mu=k_{\perp})\frac{\alpha_s^2(k_{\perp})}{\alpha_s^2(Q)}
\label{SResum}
\eea
\section{The master equation}
\label{sec:Master}

We can now go back and look at our factorized reduced density matrix and try to derive the master equation for jet evolution. From Eq.\ref{Rho}, we can write the trace over the reduced density matrix for the jet upto quadratic order in the Glauber interaction as 
\bea
\Sigma(t,\vec{q}_T) = \Sigma^{(0)}(\vec{q}_T)+ \Sigma^{(1)}(\vec{q}_T)
\label{Sig}
\eea
where we have 
\bea
\Sigma^{(1)}(\vec{q}_T) = t|C_{qq}|^2(Q)  \Sigma^{(0)}(Q,z_{c},\vec{q}_T)\otimes_{q_T}\int \frac{d^4k}{k_{\perp}^4(2\pi)^4}S^{AB}(k)\mathcal{J}_{n,\text{Med}}^{AB}(Q, z_{c},\vec{q}_T,\vec{k}_{\perp})
\eea
$C_{qq} = 8\pi \alpha_s(Q)$ is the Hard function.
We can use the RGE solutions described in the previous section to evolve medium induced soft and jet functions in rapidity and virtuality so that 
\bea
 \Sigma^{(1)}(\vec{q}_T) = t|C_{qq}|^2(Q)  \Sigma^{(0)}(Q,z_{c},\vec{q}_T)\otimes_{q_T}K_{\text{Med}}(\vec{q}_T)
\eea
with 
\bea
\label{KMed}
K_{\text{Med}}(p_{\perp}) =(N_c^2-1)\int \frac{d^2k_{\perp}}{(2\pi)^2} S_G^{\text{resum}}(k_{\perp})J^{\text{resum}}(Q, z_{c}, \vec{p}_{\perp}, k_{\perp}) 
\eea
where the resummed jet and soft functions are given in Eq. \ref{JResum} and Eq. \ref{SResum}.
Eq. \ref{Sig}  describes the evolution of the density matrix over a time scale t.  We can write the evolution equation in a suggestive form by going to impact parameter space( $\vec{b}$) which is conjugate to $\vec{q}_T$,
\bea
\Sigma(t,\vec{b}) = \Sigma^{(0)}(\vec{b})\Big[1- \frac{t}{\lambda_{\text{mfp}}(\vec{b})}\Big]
\label{Evomfp}
\eea
where 
\bea
\lambda^{-1}_{\text{mfp}}(b) = -|C_{qq}|^2K_{\text{Med}}(\vec{b})
\label{mfp}
\eea
can be thought of as  the inverse mean free path of the jet in the medium. While deriving this form ,we have assumed that t, which represents the time of evolution of the jet in the medium or equivalently the size of the medium L, is much greater than all the time scales of the jet, including the formation time of the jet $t_F \sim Q/q_T^2$. If the medium is very dilute then the mean free path(mfp) can be large compared to t in which case Eq.\ref{Evomfp} is a good enough approximation.  As we see from its definition, the mean free path is depends not only on the properties of the medium but also on the properties of the jet.  I ll explain the significance of this in the next section when I compare this formalism to previous results in literature.

The second case is when the medium is dense enough so that the mfp is comparable to the medium size then it becomes necessary to resum higher powers of $t/\lambda_{\text{mfp}}$. These higher order corrections correspond to multiple interactions of the jet with the medium. We are still working in a hierarchy when $ t \sim \lambda_{\text{mfp}} \gg t_F$ so that these multiple interactions are incoherent and hence Markovian in nature. We can therefore write a Lindblad equation by taking the limit $t \rightarrow 0$, where we assume that a single time step $t-0  = \delta t$ of the evolution is much smaller than the medium size, at the same time being much greater than $t_F$. 
We can first relate the density matrix with the cross section that we wish to compute,
\bea
 P(\vec{q}_T) \equiv \frac{d\sigma(t)}{d^2\vec{q}_T} = \mathcal{N}\frac{\Sigma(t)}{V}\nn
\eea
where $\mathcal{N}$ is a kinematical normalization factor. This leads to our master equation
\bea
\partial_tP(\vec{q}_T)(t) = P(\vec{q}_T)\otimes_{q_T} K_{\text{Med}}(\vec{q}_T)
\eea
The semi classical equation can now be solved by going to impact parameter space (conjugate to $\vec{q}_T$)
\bea
\frac{d\sigma(t)}{d^2\vec{q}_T}(e_n,t) = \int d^2\vec{b}e^{i \vec{b}\cdot \vec{q}_T}[V(\vec{b},e_n)]e^{(K_{\text{Med}}(\vec{b}))t}  
\eea
where V(b) is the Inverse Fourier Transform of the vacuum cross section. If the medium exists for a time L, then the cross section is simply evaluated by putting in t= L.

The third case is when $\lambda_{\text{mfp}}$ is much smaller than L but is comparable to the formation time of the jet $t_F$. In this case we can imagine higher order corrections of the form $t t_F/\lambda^2_{\text{mfp}} \sim t/\lambda_{\text{mfp}}$ which are now equally important and hence need to be accounted for. This correction is one in which quantum coherence is maintained over successive interactions of the jet with the medium and is well known in literature as the Landau-Pomeranchuk-Migdal(LPM) Effect. These corrections are beyond the scope of the paper but will be considered in detail in an upcoming paper \cite{Varun}. 
However, we can still give some arguments for the the factorization in this case. The important point to note is that this effect is important for high energy radiation(since $t_F \propto E$) and hence only affects the collinear jet function and not the soft function. Therefore, the factorization of the soft physics from the collinear should still hold with the Soft function remaining unchanged. This also means that the BFKL evolution which is a property of the soft function also holds and by consistency of factorization, so does the RG evolution of the jet function. I conjecture that the coherence effects will merely modify the natural rapidity scale for the jet function from its current value of Q to a smaller value involving the formation time of the jet.


%

\section{Numerical results for the medium kernel}
\label{sec:Num}

We can now look at the numerical impact of resummation for the medium kernel. 
We have from Eq. \ref{KMed}

\bea
K_{\text{Med}}(p_{\perp}) =(N_c^2-1)\int \frac{d^2k_{\perp}}{(2\pi)^2} S_G^{\text{resum}}(k_{\perp})J^{\text{resum}}(Q, z_{c}, \vec{p}_{\perp}, k_{\perp}) 
\eea
As pointed out in Eq.\ref{mfp}, we can interpret this quantity(upto a factor of $|C_{qq}|^2$) as the inverse mean free path and look at the impact of resummation at specific values of $p_{\perp}$, in this case we look at $p_{\perp} \sim T$.  

The expression for the resumed Soft function (Eq.\ref{SResum}) is straightforward to implement numerically. However, the jet function (Eq.\ref{JResum}) has a complicated solution which involves an infinite sum. For a non zero value of $p_{\perp}$, we can write the complete solution for the resummed jet function (Eq.\ref{JResum}) as 
\bea
J(\mu, \nu_f, k_{\perp}) =\frac{1}{\pi}\sum_{n=-\infty}^{\infty}\int_{1/2-i\infty}^{1/2+i\infty}\frac{d\gamma}{2\pi i} k_{\perp}^{2(\gamma-1)}e^{in\phi_k}e^{-in\phi_q}\frac{q_{Tn}^{2(\gamma^*-1)}}{q_{Tn}^2} e^{ -\frac{\alpha_s(\mu) N_c}{\pi}\chi_{n,\gamma}\ln \frac{\nu_f}{Q}} 
\eea
 Following \cite{Kovchegov:2012mbw}, can resort to a saddle point approximation in three regimes namely $k_{\perp} \ll p_{\perp}$, $k_{\perp} \sim p_{\perp}$ and $ k_{\perp} \gg p_{\perp}$ and interpolate smoothly between the three approximations. 
We consider each regime in turn ,

\begin{itemize}
\item{ $p_{\perp} \sim k_{\perp}$}\\
We can do an approximation for the $\gamma$ integral, using the saddle point approximation  when $\ln Q/\nu_f \sim \ln Q/k_{\perp}$ is large which is true in our case. Defining 
\bea
\bar \alpha_s \equiv \frac{\alpha_sN_c}{\pi}   
\eea
We can write the solution as 
\bea
 J(\mu, \nu_f, k_{\perp},\vec{q}_{T_n}) =\frac{1}{2\pi^2k_{\perp}q_{Tn}}\sqrt{\frac{\pi}{14\zeta(3)\bar \alpha_s Y}}e^{(\alpha_p-1)Y-\frac{\ln^2(k_{\perp}/q_{Tn})}{14\zeta(3)\bar \alpha_s Y}}
\eea
with 
\bea
\alpha_p-1 =\frac{4\alpha_sN_c}{\pi}\ln 2, \ \ \ \ Y= \ln \frac{Q}{\nu_f}
\eea
\item{$k_{\perp} \gg q_{Tn}$}\\
This is known as the double logarithmic approximation and we can also do a saddle point approximation here. In this case we have 
\bea
J(k_{\perp},q_{Tn})\Big|_{k_{\perp}\gg q_{Tn}} \approx  \frac{1}{2\pi^{3/2}q_{Tn}^2k_{\perp}^2}\frac{(\bar \alpha_s Y)^{1/4}}{\ln^{3/4}\left(k_{\perp}^2/q_{Tn}^2\right)}e^{2\sqrt{\bar \alpha_sY\ln(k_{\perp}^2/q_{Tn}^2)}}
\eea
The contribution from this region should be the most suppressed compared to the other two regimes.
\item{$k_{\perp} \ll q_{Tn} \sim T$}\\
Since the integral in $\gamma_I$ is symmetric under the interchange $k_{\perp} \leftrightarrow p_{\perp}$, we can again do a saddle point approximation
\bea
 J(k_{\perp}, q_{Tn})\Big|_{k_{\perp}\ll q_{Tn}} \approx \frac{1}{2\pi^{3/2}q_{Tn}^4}\frac{(\bar \alpha_s Y)^{1/4}}{\ln^{3/4}\left(q_{Tn}^2/k_{\perp}^2\right)}e^{2\sqrt{\bar{\alpha}_s Y\ln(q_{Tn}^2/k_{\perp}^2)}}
\eea
Since this is multiplied by the soft function that has a $1/k_{\perp}^2$ singularity cut-off by the gluon mass, this region will give the most contribution. 
 \end{itemize}

Once we have the approximation for the jet function in the three regimes, we can suitably interpolate between the three.
We choose a fixed perturbative $q_{Tn} \sim T~ 15$GeV with a value of $\alpha_s = 0.15$ at this scale and a hard scale $Q = 100 GeV$. The value of $m_D \sim gT$ is of the same order as the temperature for this choice of scales. This is different from the hierarchy that we have assumed where $m_D$ is much smaller than the scale of the temperature for simplicity of analysis. A separation of these scales can be achieved but only at very high temperatures which are beyond the reach of current colliders. Hence, for the present purposes, we assume an unrealistic small value of $m_D = 3GeV$ which is sufficiently separated from the scale of the temperature. The scale $m_D$ in our case therefore only serves the purpose of an IR regulator.

A more realistic analysis for the case $m_D \sim T$ would require us to keep track of all factors of $m_D$. This will not change the UV structure of the factorization but is likely to alter the rapidity evolution equation from the current conformally invariant BFKL equation to a mass dependent one. I will leave the derivation of this more phenomenologically relevant case as an added complication to be incorporated in the theoretical framework.

\begin{figure}
\centering
  \includegraphics[width=0.85\linewidth]{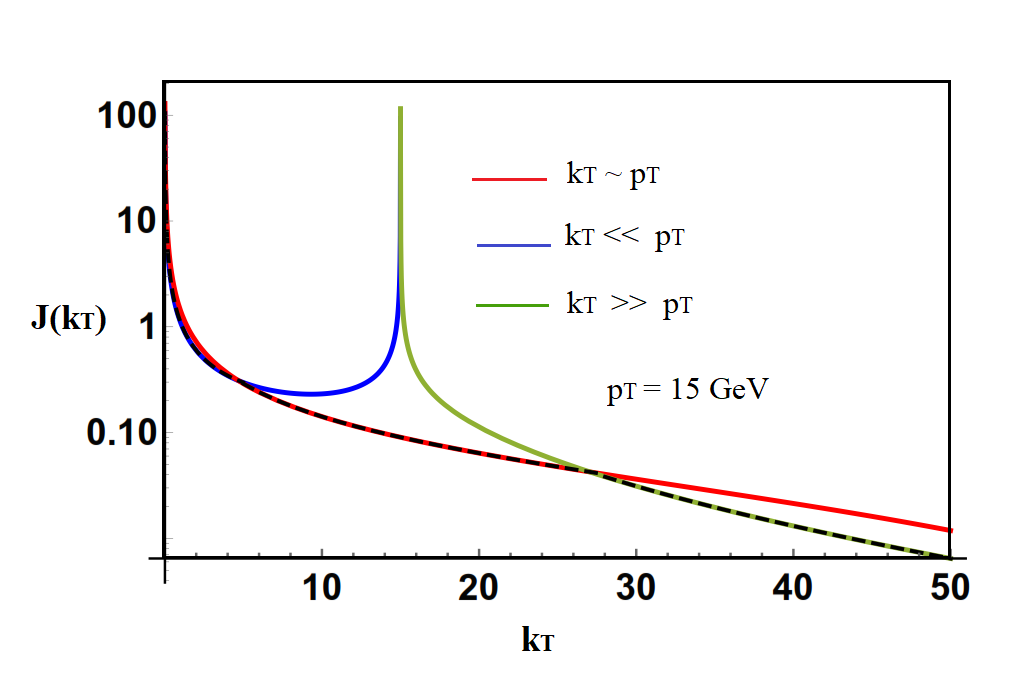}
  \caption{Approximate solutions for the resummed jet function in three regimes of $k_{\perp}$ using saddle point approximations. The black dotted curve shows a piece wise interpolation between the three approximations that we use to perform a numerical analysis.}
  \label{inter}
\end{figure}

We can now do a piece wise interpolation between our three saddle point approximations to generate a single curve for our resummed jet function valid at all $k_{\perp}$. Fig. \ref{inter} shows the approximations in the three regimes for the jet function along with an interpolating curve. 

In order to show the impact of the radiative corrections, I will give the ratio of the resummed kernel to the tree level result. 
At tree level for a non-zero $p_{\perp}$, the jet function has a simple form 
\bea
 J^{(0)}(k_{\perp}, p_{\perp})=\frac{\delta^2(\vec{p}_{\perp}-\vec{k}_{\perp})}{k_{\perp}^2} 
\eea
 while the Soft function is 
\bea
S_G^{(0)}(\vec{k}_{\perp})&=& \frac{1}{k_{\perp}^2} \int d^2p_{\perp}\int_0^{\infty} dp^+ \Bigg\{n_F\left(|\frac{p^+}{2}+\frac{p_{\perp}^2}{2}|\right)\Big[1-n_F(|\frac{p^+}{2}+\frac{(\vec{p}_{\perp}+\vec{k}_{\perp})^2}{2p^+}|)\Big]\nn\\
&+&n_F(|\frac{p^+}{2}+\frac{(\vec{p}_{\perp}+\vec{k}_{\perp})^2}{2p^+}|)\Big[1-n_F\left(|\frac{p^+}{2}+\frac{p_{\perp}^2}{2}|\right)\Big]\Bigg\}\nn\\
&\equiv& \frac{1}{k_{\perp}^2} D(k_{\perp})
\eea
so that 
\bea
K^{(0)}_{\text{Med}}(q_{\perp}) &=&\frac{N_C^2-1}{q_{\perp}^4} D(q_{\perp})
\eea
At the same time 
\bea
 K^{(1)}_{\text{Med}}(q_{\perp}) &=&(N_C^2-1)\int \frac{d^2k_{\perp}}{k_{\perp}^2+m_D^2} D(k_{\perp})\frac{\alpha^2_s(k_{\perp})}{\alpha_s(Q)^2} J^{\text{Resum}}(k_{\perp}, q_{\perp})
\eea

If we look at $D(k_{\perp})$, then this is exactly the same quantity as $\widehat{\mathcal{W}}(k_{\perp})$ which was evaluated in \cite{Vaidya:2020cyi}. From the numerical result we see that this quantity has a very mild dependence on $k_{\perp}$ and we can practically assume it to be a constant over the entire range of $k_{\perp}$ . Since we are looking at the ratio of quantities, we can therefore eliminate this factor altogether so that 
\bea
R(q_{\perp})&=& \frac{K^{(1)}_{\text{Med}}(q_{\perp}) }{K^{(0)}_{\text{Med}}(q_{\perp}) }=q_{\perp}^4\int \frac{d^2k_{\perp}}{k_{\perp}^2+m_D^2}\Bigg[\frac{\alpha_s(k_{\perp})}{\alpha_s(Q)}\Bigg]^{2}J^{\text{Resum}}(k_{\perp}, q_{\perp})
\eea

\begin{figure}
\centering
  \includegraphics[width=0.85\linewidth]{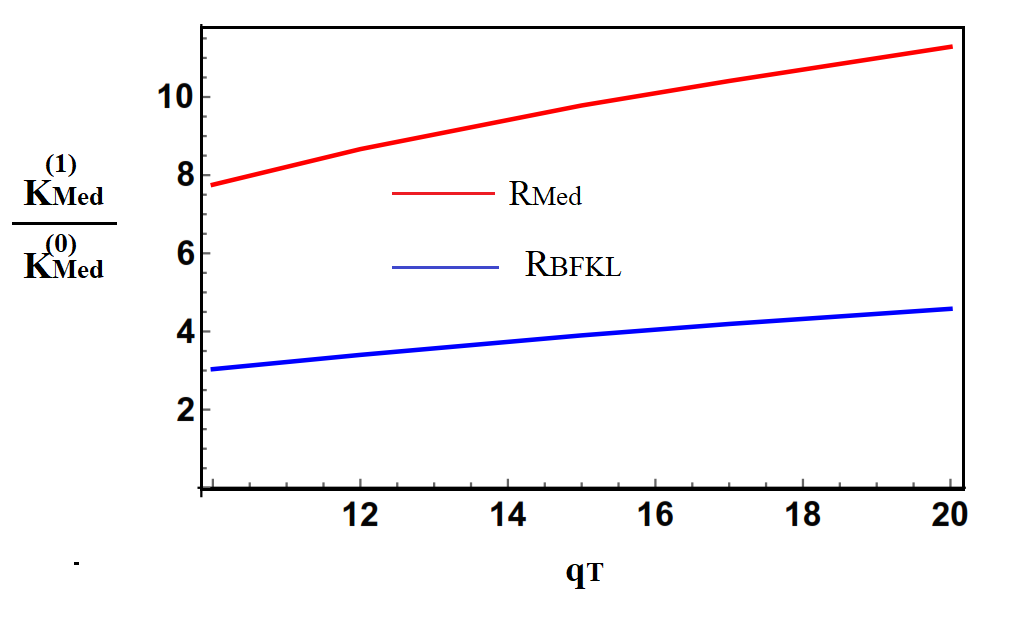}
  \caption{Enhancement for the medium Kernel due to resummed radiative corrections as a function of transverse momentum compared to the tree level result. The curve marked $R_{\text{BFKL}}$ estimates the impact of BFKL resummation in conjunction with the forward scattering singularity. The curve marked R also includes the running of the Soft function.}
  \label{RMed}
\end{figure}

From this expression we see that there two separate corrections to the medium kernel and we can explore the impact of each separately.
\begin{itemize}
\item{Soft function running}\\
We see that the correction from the running in $\mu$ is simply to replace the coupling at the hard scale Q with that at the scale $k_{\perp}$ which is sensible since the momentum transfer during the interaction of the jet with the medium is just $k_{\perp}$.  $k_{\perp}$ is free to range all the way down to the scale $m_D$, where the coupling gets stronger, at the same time this region is supported by the singularity $\sim 1/(k_{\perp}^2+m_D^2)$ which is cutoff at the scale $m_D$ so that effectively we get an enhancement $\sim \alpha^2(m_D)/\alpha^2(Q) \sim 5$ compared to the tree level result. 

\item{BFKL resummation}\\ 
The solution to the BFKL equation resums the large rapidity logarithm $ \ln Q/k_{\perp}$ along with logarithms of $q_{\perp}/k_{\perp}$ as seen in the various approximations presented earlier in this section. The singularity in the form of a factor $1/(k_{\perp}^2+m_D^2)$ from the soft function gets cut-off at the scale $q_{\perp}$ when convolved with the tree level Jet function. However, when we include radiative corrections, the BFKL resummed jet function supports this singularity all the way down to the cut-off scale $m_D$. We therefore have additional logarithmic enhancement due to the presence of this singularity. We can judge the impact of this by temporarily ignoring the effect of the Soft function running.  The curve marked $R_{\text{BFKL}}$ in Fig. \ref{RMed} presents this enhancement over a small range of values of $q_{\perp}$ about a central $ q_{\perp} \sim T = 15$ GeV. 
\end{itemize}

The same figure also shows the complete enhancement for $K_{\text{Med}}$ due to both soft and Jet resummation. This suggests that there is significant reduction in the mean free path due to enhancement of the interaction with the medium induced by radiative corrections.
In light of the discussion in the previous section, this means that due to the radiative corrections, the medium effectively appears dense which would enhance the importance of the LPM effect.

 We also have UV finite $m_D$ dependent radiative corrections due to elastic collisions presented in Section \ref{sec:Fact}. These induce logarithms $\sim \ln q_T/m_D \sim \ln 1/g$ which are sub-leading corrections compared to the large logarithm $\ln Q/k_{\perp}$ induced by BFKL evolution as well as running of the soft function which is why we have not included them in the numerical analysis. Indeed, in the phenomenologically relevant case of $m_D \sim T$, these corrections are quite small.

\section{Summary and Outlook}
\label{sec:Conclusion}

In this paper I have worked out the renormalization for the medium structure function and the medium jet function which appear in the factorization formula for jet substructure observables in heavy ion collisions. This factorization formula was derived in \cite{Vaidya:2020lih} under the following assumptions 
\begin{itemize}
\item{} The size of the medium or the time of propagation of the jet in the medium is larger than the formation time of the jet.
\item{} The medium is dilute or more precisely, the mean free path of the jet in the medium in comparable to or larger than the size of the medium.
\item{} The medium is homogeneous over the length scales probed by the jet.
\item{} The medium temperature T is high enough for the gluon mass $m_D$ to be much smaller T.
\end{itemize}
The first three of these conditions then imply that the dominant radiative corrections are those when the jet partons created in the hard interaction go on shell between successive interactions with the medium. This formalism provides a separation of physics at different scales in terms of matrix elements of manifestly gauge invariant operators. The physics of the medium is captured by the observable independent medium structure function which is defined as a gauge invariant correlator of soft currents in the background of the medium density matrix. In a sense, this function can be thought of as the "Parton Distribution Function" of the QGP that describes the probability of sourcing partons from and subsequently sinking them in the medium. The observable dependent medium induced jet function describes the evolution of high energy partons taking into account elastic collisions with the medium as well as medium induced radiation.

The radiative corrections for elastic collisions were worked out in \cite{Vaidya:2020lih} and given in section \ref{sec:Fact}. These corrections are UV finite and do not lead to any renormalization for the medium soft function and the medium induced jet function. 
In this paper, I detail the computation of the other type of corrections at one loop, namely medium induced radiation. All the divergences in the factorized functions are induced by the medium induced Bremsstrahlung in the form of UV and rapidity anomalous dimensions. In particular, the universal soft medium structure function has a rapidity anomalous dimension which leads to an Renormalization Group equation identical to the BFKL equation. The UV anomalous dimension induces the running of the QCD coupling. The medium jet function only has rapidity anomalous dimension which is equal and opposite the soft function which is a powerful check on the consistency of factorization. 
I look at the numerical impact of resummation on the mean free path of a high energy jet comparing it to the tree level result and show that the radiative corrections give significant corrections due to a combination of resummation and the forward scattering singularity which is cut-off by the medium induced gluon mass. The contribution from the medium induced radiation dominates over that from elastic collisions and effectively leads to a smaller mean free path for the jet in the medium. 

This is the first consistent factorization formalism for jet substructure observables in heavy ion collisions.  It systematically incorporates both vacuum and medium evolution of the jet and all of its leading radiative corrections are resummed using a Renormalization Group Equation.  The utility of this framework is that it tells us exactly which pieces are universal and which are process/measurement dependent by factorizing them in term of of operators.
This is still an idealized case as evidenced by the assumptions that have gone into this derivation but is an important step towards addressing a realistic scenario. In particular, we would like to consider the following phenomenologically relevant cases of increasing complexity
\begin{enumerate}
\item The medium exists for a short time comparable to or shorter than the formation time of the jet but the medium is dilute. This means that the mean free path is large compared to the medium size and only a single interaction of the jet with the medium is relevant.  In this case the jet partons wont go on-shell before they interact with the medium. This would mean a hard interaction that creates the jet and the forward scattering from the medium can have interference diagrams. Since the formation time for Soft radiation is small, we conjecture that the Soft function remains unchanged and only the jet function will get modified. However, due to consistency of factorization the rapidity RG for the jet function will also remain unchanged, albeit with a modified natural scale for the rapidity. 

\item A more complicated case would be that the mean free path can be comparable to or smaller than the formation time of the jet which in turn is comparable to the size of the medium. This would mean that apart from hard-medium interaction interference,  multiple interactions of the jet with the medium are important and there is quantum interference between successive interactions(LPM effect). Again, this will only impact the jet function which by RG consistency should still have a BFKL type evolution. This entails defining and computing jet functions at higher orders in the Glauber Hamiltonian and summing the series leading to possibly an AMY (\cite{Arnold:2002ja}) type formalism, albeit with a BFKL resummation implemented at each order.

\item The Debye mass scale $m_D$ is comparable to the scale of the temperature. This is relatively straightforward  to address in that we need to keep the complete $m_D$ dependence which will likely lead to a massive version of the BFKL equation.

\end{enumerate}
Many of these interesting open questions will be addressed in an upcoming paper \cite{Varun}.


\acknowledgments
I thank Iain Stewart and Yuri Kovchegov for useful discussions on several conceptual aspects of this project. This work is supported by the Office of Nuclear Physics of the U.S. Department of Energy under Contract DE-SC0011090 and Department of Physics, Massachusetts Institute of Technology.

\end{document}